\documentclass[a4paper,11pt]{article}

\usepackage[left=2.5cm,right=2.5cm,top=2.5cm,bottom=2.5cm]{geometry}
\usepackage{graphicx,amssymb,amsmath,amsthm,mathrsfs,setspace,subcaption,cite,authblk,float}
\usepackage[all]{xy}

\usepackage[colorlinks=true,bookmarks=true]{hyperref}
\usepackage{breakurl} 

\usepackage{mathtools}

\DeclarePairedDelimiterX\braket[2]{\langle}{\rangle}{#1 \delimsize\vert #2}

\theoremstyle{definition}
\newtheorem{mydef}{Definition}
\newtheorem{mythm}{Theorem}
\newtheorem{myex}{Example}

\hypersetup{citecolor=blue}
\newcommand{\dif}{\mathrm{d}}
\newcommand{\Eqref}[1]{(\ref{#1})}
\newcommand{\half}{\frac{1}{2}}

\newcommand{\brac}[1]{\left(#1 \right)}
\newcommand{\sbrac}[1]{\left[#1\right]}

\newcommand{\Log}{\mathscr{L}\mathrm{og}_{\mathrm{C}}}
\newcommand{\di}{\displaystyle}

\numberwithin{equation}{section}
 
\begin{document}

\title{Light-ring pairs from $A$-discriminantal varieties}

\author[1]{Yen-Kheng Lim\footnote{Email: yenkheng.lim@gmail.com, yenkheng.lim@xmu.edu.my}}

\author[2]{Mounir Nisse\footnote{Email: mounir.nisse@gmail.com, mounir.nisse@xmu.edu.my}}

\affil[1]{\normalsize{\textit{Department of Physics, Xiamen University Malaysia, 43900 Sepang, Malaysia}}}

\affil[2]{\normalsize{\textit{Department of Mathematics, Xiamen University Malaysia, 43900 Sepang, Malaysia}}}

\date{\normalsize{\today}}
\maketitle 

\renewcommand\Authands{ and }
\begin{abstract}
 When geodesic equations are formulated in terms of an effective potential $\mathcal{U}$, circular orbits are characterised by $\mathcal{U}=\partial_a\mathcal{U}=0$. In this paper we consider the case where $\mathcal{U}$ is an algebraic function. Then the condition for circular orbits defines an $A$-discriminantal variety. A theorem by Rojas and Rusek, suitably interpreted in the context of effective potentials, gives a precise criteria for certain types of spacetimes to contain at most two branches of light rings (null circular orbits), where one is stable and the other one unstable. We identify a few classes of static, spherically symmetric spacetimes for which these two branches occur.
\end{abstract}

\section{Introduction} \label{intro}

Recent breakthroughs in observations of gravitational phenomena has made it increasingly relevant to understand the gravitational field and the motion of light in the presence of strong gravity. For instance, the spherical photon surfaces are related to the understanding of the ringdown phase of black-hole mergers, as well as the optical shadow of a black hole. The former has been observed via gravitational wave observations \cite{Abbott:2016blz} and the latter was observed for the black hole at the centre of the galaxy M87 \cite{Akiyama:2019cqa}.  

A question related to these issues is whether the shadow and ringdown phenomena can be conclusively identified with black holes, or whether these signals could  be mimicked by compact objects with no event horizon. Cunha, Berti, and Herdeiro have shown in Refs.~\cite{Cunha:2017qtt} that stationary, compact (non-black hole) objects formed from incomplete gravitational collapse have light rings which form in pairs. Hod in Ref.~\cite{Hod:2017zpi} provided an important supplement to their result. On the other hand, it was also shown in \cite{Cunha:2020azh} that generic black holes must have at least one unstable light ring. Further related results have been provided by Guo and Gao \cite{Guo:2020qwk}. The presence of stable photon orbits may signal an instability of the spacetime \cite{Keir:2014oka,Cardoso:2014sna}. More recently, it was proven that any four-dimensional, stationary, axi-symmetric spacetime with an ergoregion must have at least one light ring outside the ergoregion \cite{Ghosh:2021txu}. These results hold for horizonless objects with an ergoregion such as in \cite{Chirenti:2008pf}. Therefore there is an increasing body of work showing that very compact horizonless objects have a stable light rings, while black holes are often characterised by the presence of unstable light rings. (It should be noted that there are some examples of black holes with a stable light-ring exterior to their horizon; for instance, in \cite{Cunha:2016bjh}.)

In this paper, we approach these questions from the direction of algebraic geometry. In terms of an effective potential $\mathcal{U}$, circular time-like or null geodesics are determined by the condition $\mathcal{U}=\partial_a\mathcal{U}=0$. If $\mathcal{U}$ happens to be an algebraic function, this condition is equivalent to $\mathcal{U}$ having a vanishing discriminant. Therefore the circular geodesics can be represented as an algebraic variety in the parameter space, known as the \emph{$A$-discriminant}. 

We explore the following in the main body of this paper: A theorem by Rojas and Rusek \cite{Rusekthesis,Rusek2016} states that if $\mathcal{U}$ is a polynomial of $n$ variables and is a sum of $n+3$ monomials, the contour of the amoeba corresponding to the $A$-discriminant can have at most $n$ cusps. In the context of the geodesic effective potential, the cusps represent circular orbits of marginal stability, thus separating branches of stable and unstable orbits. Therefore this theorem provides a constraint on the kinds of circular orbits that can exist in a given spacetime, provided that the effective potential satisfies the afore-mentioned conditions.

In the case of static, spherically-symmetric spacetimes, the corresponding effective potential $\mathcal{U}$ typically takes the form of a univariate ($n=1$) polynomial. Therefore, Rojas and Rusek's theorem is appliable if $\mathcal{U}$ consists of a sum of $n+3=4$ monomials, for instance in the form 
\begin{align}
 \mathcal{U}\propto ax^m+bx^l+x^k+1, \label{Uform}
\end{align}
which says that the $A$-discriminant can have at most $n=1$ cusp. Consequently this means these spacetimes can have at most two branches of circular orbits, one stable and the other unstable. 

While this condition may appear highly restrictive, there are several known spacetimes that do satisfy this condition, in particular, the effective potentials of time-like geodesics in Schwarzschild spacetime, and of null geodesics in the Hayward and Reissner--Nordstr\"{o}m spacetimes. These examples are worked out explicitly in this paper. Furthermore, without assuming the field equations of any theory, we consider generic spacetimes whose null geodesic effective potential takes the form \Eqref{Uform}. We show in this paper that the stable branch of light rings only occur in the case of horizonless (non-black hole) compact objects, thus lending support to the arguments of \cite{Cunha:2017qtt,Cunha:2020azh,Hod:2017zpi,Guo:2020qwk} from the point of view of $A$-discriminants. We also show that the results of Rojas and Rusek's theorem continue to hold when the effective potential is perturbed with additional monomials beyond $n+3$.

The rest of this paper is organised as follows. In Sec.~\ref{sec_geodesics} we review the geodesic equations in spacetimes with a certain number of Killing vectors and define the effective potentials. The theory of $A$-discriminantal varieties, along with Rojas and Rusek's theorem are reviewed in Sec.~\ref{sec_Adiscrim}, where we will also explain how the theorem applies to circular geodesics. As an demonstrative example of the formalism, we obtain the circular time-like orbits around the Schwarzschild black hole in Sec.~\ref{sec_Sch} using $A$-discriminants. Subsequently in Sec.~\ref{sec_classes}, we find generic forms of static, spherically symmetric spacetimes which satisfy the conditions of Rojas and Rusek's theorem, giving us spacetimes with light-ring pairs. Some known spacetimes that satisfy these conditions are worked out explicitly in Sec.~\ref{sec_examples}. In Sec.~\ref{sec_beyond} we consider a case where the spacetime is not spherically symmetric, and also show how Rojas and Rusek's theorem continue to hold beyond $n+3$ monomials. Conclusions and closing remarks are given in Sec.~\ref{sec_conclusion}.

%
%
%
%

\section{Geodesics} \label{sec_geodesics}


We start by establishing some general properties of the background spacetime. We consider a $(D=p+n)$-dimensional spacetime manifold $\mathcal{M}$ that possesses a time-like Killing vector field $\xi_{(0)}$ which generates $\mathbb{R}$-isometries. We further suppose $\mathcal{M}$ has another $(p-1)$ space-like Killing vector fields $\xi_{(1)},\ldots,\xi_{(p-1)}$, each generating a $U(1)$ isometry. Altogether, the full set of $p$ commuting Killing vector fields
\begin{align}
 \{\xi_{(0)},\xi_{(1)},\ldots,\xi_{(p-1)}\} \label{Killings}
\end{align}
generates the isometry group $\mathbb{R}\times U(1)^{p-1}$. Spacetimes of this form have been studied in detail in Refs.~\cite{Emparan:2001wk,Harmark:2004rm,Harmark:2005vn,Harmark:2009dh,Chen:2010zu}. For such a spacetime, we can always find a coordinate system $(\sigma^0,\sigma^1,\ldots\sigma^{p-1},x^1,\ldots,x^n)$ such that each $\sigma^M$ is a coordinate Killing direction, that is,
\begin{align}
 \xi_{(M)}=\frac{\partial}{\partial\sigma^M},\quad M=0,1,\ldots,p-1.
\end{align} 
It was further shown that if $\xi_{(0)}^{[\mu_0}\xi_{(1)}^{\mu_1}\cdots\xi_{(p-1)}^{\mu_{p-1}}\nabla^\nu\xi^{\lambda]}_{(M)}=0$ holds for all $M=0,1,\ldots,p-1$, the metric on $M$ can be written in the form \cite{Harmark:2009dh}
\begin{align}
 \dif s^2&=g_{\mu\nu}\dif y^\mu\dif y^\nu=G_{MN}\dif\sigma^M\dif\sigma^N+\bar{g}_{ab}\dif x^a\dif x^b,
\end{align}
where the metric components $G_{MN}$ and $\bar{g}_{ab}$ depend on coordinates $x^a$ only. 

At this point, it is worth noting some subtle differences in notions which are variously called \emph{photon surfaces}, \emph{photon spheres}, and \emph{light rings} in the literature. Intuitively, the general phenomena of interest is where light/photons are in bound orbit around a gravitating body. In Refs.~\cite{Claudel:2000yi,Perlick:2005,Cederbaum:2014gva,Cederbaum:2015fra} the authors consider four-dimensional static spacetimes and defined a photon surface as a co-dimension 1 submanifold $\mathcal{P}\subset\mathcal{M}$, where any null geodesic initially tangent to $\mathcal{P}$ remains tangent throughout the entire geodesic. This has been extended to higher dimensions in \cite{Gallo:2015bda}. The authors of \cite{Gibbons:2016isj,Cvetic:2016bxi} use a similar definition, first by defining the \emph{optical metric} of $\mathcal{M}$, then the photon surface is a totally geodesic submanifold of the optical manifold defined by the optical metric. Geometrically, the surface need not be spherical; the particular case where the photon surface is a \emph{photon sphere} is defined to be the case where the lapse function $g_{tt}=g(\xi_{(0)},\xi_{(0)})\equiv-N^2$ is constant over $\mathcal{P}$. This definition is coordinate independent and makes no reference to any isometries apart from the time-like Killing vector $\xi_{(0)}$.

On the other hand, if all the Killing isometries are invoked to (partially) separate the Hamilton--Jacobi equations with their associated conserved quantities, the equations of motion can be cast in terms of an effective potential $\mathcal{U}$ which is a function that depends on $x^a$ only. Then, the \emph{light rings} are null geodesics which can be obtained from $\mathcal{U}=\partial_a\mathcal{U}=0$. This `defnition' is less precise as it depends on how one defines the effective potential. In fact, this can be clarified by considering the following situation in which we can have two ways of writing down the potential. If the equations of motion possess hidden symmetries \cite{Frolov:2017kze} the equations can be further separated and we have various effective potentials $\mathcal{U}_i$ which are functions depending on distinct subsets of coordinates $\{x^a\}$.

For example, in Boyer--Lindquist coordinates $(t,r,\theta,\phi)$, the Kerr spacetime is axi-symmetric with two Killing vectors $\xi_{(0)}=\frac{\partial}{\partial t}$ and $\xi_{(1)}=\frac{\partial}{\partial\phi}$. These give rise to conservation of energy $E$ and anglar momentum $L$, and the effective potential can be easily written as $\mathcal{U}(r,\theta)$. However it is well known that Kerr geodesics possess a hidden symmetry \cite{Carter:1968rr} so that the equations of motion can be completely separated into $\dot{r}=R(r)$ and $\dot{\theta}=\Theta(\theta)$. Then, $R(r)$ and $\Theta(r)$ can be interpreted as the (negative of the) effective potentials in their respective directions. In this context, the \emph{photon sphere} \cite{Nemiroff:1993he,Teo:2003} of Kerr geometry refers to the case $R(r)=R'(r)=0$ while $\theta$ is allowed to vary. On the other hand, taking the definition $\mathcal{U}(r,\theta)=\partial_a\mathcal{U}(r,\theta)=0$ means both $r$ and $\theta$ are constant throughout the motion, so the geodesics are simply circles. Hence, in this paper we refer to this situation as the \emph{light ring}.

We now turn our attention to setting up the geodesic equations explicitly. The following discussion covers both null and time-like geodesics. In either case, they are described by affinely parametrised curves of the form
\begin{align}
 y^\mu(\tau)=\brac{\sigma^M(\tau),x^a(\tau)}.
\end{align}
We take $\tau$ to be an appropriate affine parametrisation such that $\dot{y}^\mu\dot{y}_\mu=\epsilon$, where over-dots denote derivatives with respect to $\tau$ and $\epsilon=-1$ for time-like geodesics and $\epsilon=0$ for null geodesics. 

To derive the equations for geodesic motion, we begin with the Lagrangian 
\begin{align}
 \mathcal{L}=\half\brac{G_{MN}\dot{\sigma}^M\dot{\sigma}^N+\bar{g}_{ab}\dot{x}^a\dot{x}^b}.
\end{align}
From the Lagrangian, we define the canonical momenta,
\begin{align}
 p_i&=\frac{\partial\mathcal{L}}{\partial\dot{x}^a}=\bar{g}_{ab}\dot{x}^b,\quad
 P_M=\frac{\partial\mathcal{L}}{\partial\dot{\sigma}^M}=G_{MN}\dot{\sigma}^N.
\end{align}
As each $\sigma^M$ is adapted to the Killing directions, they become cyclic coordinates of the Lagrangian, and hence, $P_M$ are constants of motion.

The Hamiltonian is obtained from the Legendre transform of the Lagrangian, $\mathcal{H}(p,y)=p_\mu\dot{y}^\mu-\mathcal{L}$. Explicitly, we have 
\begin{align}
 \mathcal{H}=\half\brac{G^{MN}P_MP_N+\bar{g}^{ab}p_ap_b},
\end{align}
where $G^{MN}$ and $\bar{g}^{ab}$ are the inverse of $G_{MN}$ and $\bar{g}_{ab}$, respectively. Furthermore, the condition $\dot{y}^\mu\dot{y}_\mu=\epsilon$ leads to the constraint
\begin{align}
 \bar{g}^{ab}p_ap_b+\mathcal{U}=0,
 \end{align}
where we have defined the effective potential 
\begin{align}
 \mathcal{U}=G^{MN}P_MP_N-\epsilon.
\end{align}
As mentioned above, $\sigma^M$ are cyclic coordinates and their conjugate momenta $P_M$ are constants of motion. This means $\mathcal{U}$ is a function of $x^a=(x^1,\ldots,x^n)$ only.

We are primarily interested in geodesics where the coordinates $x^a$ are constant. The Hamiltonian equations for these coordinates are 
\begin{subequations}
\begin{align}
 \frac{\partial\mathcal{H}}{\partial x^a}=\half\brac{\partial_a G^{MN}}P_MP_N+\half\brac{\partial_a\bar{g}^{cd}}p_cp_d&=-\dot{p}_a,\\
 \frac{\partial\mathcal{H}}{\partial p_a}=\bar{g}^{ab}p_b&=\dot{x}^a.
\end{align}
\end{subequations}
Assuming $\bar{g}^{ab}$ is non-degenerate, the constancy of $x^a$ requires $p_a=\dot{p}_a=0$. This, in turn leads to
\begin{subequations} \label{circular_cond}
\begin{align}
 \brac{\partial_a G^{MN}}P_MP_N=\partial_a\mathcal{U}&=0,\\
           \mathcal{U}&=0.
\end{align}
\end{subequations}
In the context of the preceding discussions in this section, it may be apt to refer to geodesics satisfying \Eqref{circular_cond} as \emph{circular geodesics}, as they will be ring shaped (circles) in the case of four-dimensional, axi-symmetric spacetimes. In particular, for null geodesics ($\epsilon=0$), these are the \emph{light rings}.


Suppose that the effective potential is now cast in the form
\begin{align}
 -\mathcal{U}=h(x)F(x), \label{prime_eqn2}
\end{align}
where $x=(x^1,\ldots,x^n)$ and $h(x)$ will be some function strictly positive or strictly negative in the domain of consideration such that Eq.~\Eqref{circular_cond} is equivalent to 
\begin{align}
 F(x)=\partial_a F(x)=0. \label{circular_cond2}
\end{align}
In practice, the function $h(x)$ is simply a result of rescaling/rearranging of factors of $-\mathcal{U}$ such that $F(x)$ takes a convenient form. (It will be clear in the examples that follow.) Stable circular orbits/light rings can be characterised by whether $F(x)$ is a local maximum or local minimum, along with noting the sign of $h(x)$, which ultimately determines the extremum properties of $\mathcal{U}$.

At this stage, we observe that if $F(x)$ is a polynomial, Eq.~\Eqref{circular_cond2} is precisely the defining conditions for $F(x)$ to have a vanishing discriminant. In other words, the condition \Eqref{circular_cond2} describes an $A$-discriminantal variety. In this case, the theory of $A$-discriminantal varieties can potentially provide an insight to circular geodesics. This is reviewed in the following section.

\section{\texorpdfstring{$A$}{A}-discriminants} \label{sec_Adiscrim}

\subsection{Brief review of the \texorpdfstring{$A$}{A}-discriminants and the Horn--Kapranov uniformisation}

A common setting for the theory of $A$-discriminants is the space of complex numbers. Here, we denote an $n$-tuple of non-zero complex numbers by $z=(z_1,\ldots,z_n)\in(\mathbb{C}^*)^n$, where $\mathbb{C}^*=\mathbb{C}\setminus\{0\}$ is the complex plane with the zero point removed. Let $A$ be a finite configuration of points $\{\alpha_{1},\ldots,\alpha_{N} \}\subset\mathbb{Z}^n$ where $\alpha_{i}=\brac{\alpha_{i}^1,\ldots,\alpha_{i}^n}\in\mathbb{Z}^n$ for each $i$. For a given $A$ we have the corresponding family $(\mathbb{C}^*)^A$ of Laurent polynomials of $N$ terms,
\begin{align}
 f(z)=\sum_{i=1}^Na_i z^{\alpha_{i}},\quad\mbox{ where }\quad z^{\alpha_{i}}=z_1^{\alpha_{i}^1}\cdots z_n^{\alpha_{i}^n},
\end{align}
with exponent vectors from $A$. The polynomial $f(z)$ is identified with the point $f=\brac{a_1,\ldots,a_N}\in(\mathbb{C}^*)^N$. If we denote by $Z_f$ the zero locus of $f$, i.e., the set 
\begin{align*}
 Z_f=\left\{z\in(\mathbb{C}^*)^n \;|\; f(z)=0 \right\}.
\end{align*}
Then the set of coefficient vectors $(a_1,\ldots,a_N)$, for which the hypersurface $Z_f$ is not a smooth manifold coincides with the zero locus of an irreducible polynomial $D_A\in\mathbb{Z}[a_1,\ldots,a_N]$. The notion of $A$-discriminants was introduced by Gelfand, Kapranov, and Zelevinsky (see\cite{GKZ-94}) is based on the idea that one should study the whole family $(\mathbb{C}^*)^A$ rather than a single polynomial $f$.

Let us give a precise description of the $A$-discriminant. A point $z\in(\mathbb{C}^*)^n$ is said to be a critical point of a polynomial $f$ if it is a solution of the system of equations 
\begin{align}
 \frac{\partial f}{\partial z_i}(z)=0\quad\mbox{for all}\quad i=1,\ldots,n.
\end{align}
The $A$-discriminant $D_A(f)$ is, by definition, an irreducible polynomial which vanishes if and only if $f$ has a singular point in $(\mathbb{C}^*)^n$, i.e., a critical point $z$ with $f(z)=0$. Moreover, this $A$-discriminant $D_A$ is uniquely determined up to sign, provided its coefficients are taken to be relatively prime. Also, a $\mathbb{Z}$-affine isomorphism of the point configuration $A$ leaves the $A$-discriminant invariant. By a slight abuse of notation, we denote by the same letter $A$ its associated $(1+n)\times N$ matrix 
\begin{align}
 A=\left(\begin{array}{ccc}
          1 & \cdots & 1 \\
          \alpha_1 & \cdots & \alpha_N
         \end{array}\right)
         =\left(\begin{array}{cccc}
                 1 & 1 & \cdots & 1 \\
                 \alpha_1^1 & \alpha_2^1 & \cdots & \alpha^1_N \\
                 \vdots & \vdots & \ddots & \vdots \\
                 \alpha_1^n & \alpha_2^n & \cdots & \alpha_N^n
                \end{array}\right)
\label{A_matrix}
\end{align}
where each $\alpha_i\in\mathbb{Z}^n$ is viewed as column $n$-vectors for $i=1,\ldots,N$. The configuration $A$ gives rise to a lattice $\Lambda(A)=\mathbb{Z}A\subset\mathbb{Z}^n$ of index $\mathrm{ind}(A)=[\mathbb{Z}^n:\Lambda(A)]$. 

We make the following assumptions about the matrix $A$: (i) $\mathrm{rank}(A)=1+n$, and (ii) the maximal minors of $A$ are relatively prime. This means that the columns of $A$ generate the full lattice $\mathbb{Z}^{n+1}$. 

Since every row vector of the matrix $A$ corresponds to a (quasi-)homogeneity of the $A$-discriminant, then the $A$-discriminant $D_A$ can be considered as a function of only $m=N-n-1$ essential variables instead of $N$ variables. To dehomogenise the $A$-discriminant, one can choose a Gale transform of $A$. In other words, we can choose an integer $(N\times m)$-matrix $B$, whose column vectors form a $\mathbb{Z}$-basis for the kernel of the linear map represented by the matrix $A$. Therefore, the row vectors $b_1,\ldots,b_N$ of $B$ constitute a point configuration in $\mathbb{Z}^m$ called a \emph{Gale transform} of the original configuration $A$. 

To be more precise, we identify $B$ with the matrix $(b_1,\ldots,b_N)^{\mathrm{T}}$. Explicitly, the entries of $B$ are 
\begin{align}
 B=\left(\begin{array}{cccc}
          b_1^1 & b_1^2 & \cdots & b_1^m \\
          \vdots & \vdots & \ddots & \vdots \\
          b_N^1 & b_N^2 & \cdots & b_N^m
         \end{array}\right).
\end{align}
This $B$ is called a \emph{Gale dual} of $A$ if the columns of $B$ span the kernel of $A$. In other words, $B$ is a Gale dual of $A$ if the matrix $B$ has maximal rank and 
\begin{align}
 AB=0. \label{B_def}
\end{align}
Then we note that Eq.~\Eqref{B_def} implies
\begin{subequations} \label{Au_rowsum}
\begin{align}
 \sum_{j=1}^Nb_i^j&=0,\quad i=1,\ldots,m, \label{Au_rowsum1}\\
 \sum_{j=1}^N \alpha_i^j b_k^j&=0,\quad i,k=1,\ldots,m.\label{Au_rowsum2}
\end{align}
\end{subequations}
%
%
Also, this means that the column vectors of $B$ can be used to produce inhomogeneous coordinates for $D_A$. In fact, the reduction to only $m$ variables corresponds to a choice of Gale dual $B$ of $A$, and the induced projection $\pi_B:(\mathbb{C}^*)^A\rightarrow(\mathbb{C}^*)^m$. Explicitly, using coordinates, we have 
\begin{align}
 x_j=a_1^{b^j_1}a_2^{b^j_2}\cdots a_N^{b^j_N},\quad j=1,\ldots,m.
\end{align}
Therefore, there exists a Laurent monomial $M(f)$ in the original $a$-variables, and a polynomial $D_B(x)$ such that
\begin{align}
 D_A(a_1,\ldots,a_N)=M(f)D_B(x_1,\ldots,x_m).
\end{align}
where $D_B$ is called the corresponding \emph{reduced} $A$-discriminant.

Now, we can equally start the theory from a $B$-matrix whose row vectors sum up to zero, and then take a Gale transform $A$ of the form \Eqref{A_matrix} which will be uniquely determined up to a $\mathbb{Z}$-affine isomorphism. 

Kapranov proved \cite{Kapranov:1991} that the zero locus of the reduced $A$-discriminant $D_B$ is the image of the projective space by a birational map $\Psi$, the so-called Horn--Kapranov parametrisation \cite{Horn:1889,Kapranov:1991}. More precisely, $\Psi$ is a birational equivalence whose inverse is the logarithmic Gauss mapping. Let us recall the definition of the logarithmic Gauss mapping which we denote by $\gamma$. First, $\gamma$ is defined on the smooth part of a complex hypersurface $V\subset(\mathbb{C}^*)^m$ with defining polynomial $f$ as follows:
\begin{align}
 \gamma:\mathrm{reg}(V)&\rightarrow\mathbb{CP}^{m-1},\nonumber\\
              x&\mapsto\sbrac{x_1\frac{\partial f}{\partial x_1}(x):\ldots : x_m\frac{\partial f}{\partial x_m}(x)},
\end{align}
where $\mathrm{reg}(V)$ denotes the smooth part of $V$. Geometrically, given a smooth point $x_0\in V$, we choose a local holomorphic branch around the point $x_0$ of the complex logarithmic map $\Log$,
\begin{align}
 x_0=\brac{x_{01},\ldots x_{0m}}\mapsto\brac{\Log(x_{01}),\ldots,\Log(x_{0m})}.
\end{align}
Then, $\gamma(x_0)\in\mathbb{CP}^{m-1}$ is the complex normal direction to $\Log(V)$ at $\Log(x_0)$. It was proved by Kapranov \cite{Kapranov:1991} that if the hypersurface $V$ is defined by a reduced $A$-discriminant $D_B$, then the logarithmic Gauss mapping is birational, with inverse the rational mapping $\Psi$ called Horn--Kapranov parametrisation defined as follows.

\begin{mydef}
 The Horn--Kapranov parametrisation of the discriminant hypersurface defined as the zero locus of $D_B$ is the rational mapping $\Psi:\mathbb{CP}^{m-1}\rightarrow(\mathbb{C}^*)^m$ given by 
 \begin{align}
  \Psi[t_1:\ldots:t_m]=\brac{\prod_{j=1}^N\langle b_j,t\rangle^{b_{j}^1},\prod_{j=1}^N\langle b_j,t\rangle^{b_{j}^2},\ldots,\prod_{j=1}^N\langle b_j,t\rangle^{b_{j}^m}}. \label{psi_def}
 \end{align}
\end{mydef}
Let us illustrate this concept in a few classical and easy examples. 

\begin{myex}
 Consider the trivial $A$-matrix $A=(1\;1\;1)$. This means that $n=0$. We can define the corresponding $A$-discriminant to be the linear map $D_A=a_1+a_2+a_3$, and a $B$-matrix for this case is given by 
 \begin{align}
  B=\left(\begin{array}{rr}
           -1 & -1 \\ 1 & 0 \\ 0 & 1
          \end{array}\right).
 \end{align}
 Then, we can set $x_1=a_1^{-1}a_2$ and $x_2=a_1^{-1}a_3$, and then $D_A(a)=a_1D_B(x)$ with $D_B(x)=1+x_1+x_2$. Hence, the complex line can be seen as the zero locus of a reduced $A$-discriminant. More precisely, the Horn--Kapranov mapping in this case is given by 
 \begin{align}
  \Psi[1:t]=\brac{-\frac{1}{1+t},-\frac{t}{1+t}},
 \end{align}
 which parametrises the complex line with defining polynoimal $D_B(x_1,x_2)=1+x_1+x_2$. In fact, set $x_1=-\frac{1}{1+t}$ and $x_2=-\frac{t}{1+t}$ then we get $x_1+x_2=-1$.
\end{myex}

\begin{myex}
 Let $n=1$ and $A$ be the following configuration:
 \begin{align}
  A=\left(\begin{array}{cccc}
           1 & 1 & 1 & 1 \\ 0 & 1 & 2 & 3
          \end{array}\right).
 \end{align}
 The $A$-discriminant in this case is given by
 \begin{align}
  D_A(a)=27a_1^2a_4^2+4a_1a_3^3-18a_1a_2a_3a_4-a_2^2a_3^2.
 \end{align}
 This is the classical cubic discriminant that vanishes precisely when the degree-3 polynomial equation $a_1+a_2x+a_3x^2+a_4x^3=0$ has a multiple root. The Gale dual of $A$ is given by the $B$-matrix 
 \begin{align}
  B=\left(\begin{array}{rr}
           1 & 0 \\ 0 & 1 \\ -3 & -2 \\ 2 & 1
          \end{array}\right).
 \end{align}
 Let $x_1=a_1a_3^{-3}a_4^2$ and $x_2=a_2a_3^{-2}a_4$, and then we get $D_A(a)=a_3^6a_4^{-2}D_B(x)$ with the reduced $A$-discriminant given by $D_B(x)=27x_1^2+4x_1+4x_2^3-18x_1x_2-x_2^2$. Using Horn--Kapranov in our case, we get
 \begin{align}
  x_1=-\frac{(2+t)^2}{(3+2t)^3},\quad\mbox{ and }\quad x_2=\frac{t(2+t)}{(3+2t)^2}.
 \end{align}
 This gives a parametrisation of the reduced cubic discriminant curve 
 \begin{align*}
  27x_2+4x_1+4x_2^3-18x_1x_2-x_2^2=0.
 \end{align*}

\end{myex}

\begin{myex}
 Let $n=2$ and let $A$ be the following configuration
 \begin{align}
  A=\left(\begin{array}{ccccc}
           1 & 1 & 1 & 1 & 1 \\ 0 & 2 & 0 & 1 & 2 \\ 0 & 0 & 3 & 3 & 2
          \end{array}\right).
 \end{align}
 In this case one can choose the Gale dual $B$-matrix to be 
 \begin{align}
  B&=\left(\begin{array}{rr}
            1 & 2 \\ -1 & -3 \\ -2 & -2 \\ 2 & 0 \\ 0 & 3
           \end{array}\right).
 \end{align}
 The variables $x_1$ and $x_2$ are as follows: $x_1=a_1a_2^{-1}a_3^{-2}a_4^2$ and $x_2=a_1^{2}a_2^{-3}a_3^{-2}a_5^{3}$ and the reduced $A$-discriminant $D_B$ given by 
 \begin{align}
  D_B(x)&=729x_2-1280x_1^2 + 2187 x_1^3 + 2187x_1^4 + 729 x_1^5 + 1728x_2\nonumber\\
   &\quad -4752x_1x_2 + 5400 x_1^2x_2 - 1404x_1^3x_2 - 864x_1^4x_2 + 3456x_2^2 - 5616x_1x_2^2\nonumber\\
   &\quad +576x_1^2x_2^2 + 256x_1^3x_2^2 + 1728x_2^3
 \end{align}
 Using the Horn--Kapranov parametrization, we get:
 \begin{align}
  x_1=-\frac{(1+2t)}{(1+3t)(1+t)^2},\quad\mbox{and}\quad x_2=-\frac{9t^3(1+2t)^2}{4(1+3t)^3(1+t)^2}.
 \end{align}
 We can verify  that it does satisfy $D_B(x_1,x_2)=0$, where $D_B$ is the reduced $A$-discriminant as above.

\end{myex}

As $D_B$ is an algebraic variety, we may define its \emph{amoeba} \cite{GKZ-94} by the log map $\mathrm{Log}:(\mathbb{C}^*)^{k-1}\rightarrow\mathbb{R}^{k-1}$ defined by 
\begin{align}
 \mathrm{Log}(z_1,\ldots,z_n)=\brac{\log|z_1|,\ldots,\log|z_n|}.
\end{align}
The image of $D_B$ under this map is the amoeba, denoted by $\mathcal{A}$. Denoting $\Phi=\mathrm{Log}\circ\psi$, a point $\lambda\in\mathbb{CP}^{m-1}$ maps to a point in $\mathcal{A}$ that is given in the explicit formula by
\begin{align}
 \Phi(\lambda)&=\brac{\varphi_1,\ldots,\varphi_{m}}, \label{varphi_def}
\end{align}
where the components are
\begin{align}
 \varphi_i=\log\left|\prod_{j=1}^{N}\langle b_j,\lambda\rangle^{b_j^{i}}\right|=\sum_{j=1}^{N}b^{i}_j\log\left|\langle b_j,\lambda\rangle\right|,\quad i=1,\ldots,m.
\end{align}

Our discussion so far can be summarised in the following diagram:
\begin{align*}
\xymatrix{      \mathbb{CP}^{m-1} \ar[dr]_{\Phi} \ar[r]^{\Psi}& D_B \ar[d]^{\mathrm{Log}} \\
      & \mathcal{A}
}
\end{align*}
Here, the reduced $A$-discriminant $D_B$ lives in $(\mathbb{C}^*)^{m}$, and its amoeba $\mathcal{A}$ lives in $\mathbb{R}^{m}$. Theorem 2.1 of \cite{Kapranov:1991} and Lemma 3 of \cite{Mikhalkin:2000} combines to tell us that the image of the map $\Phi$, when restricted to $\mathbb{RP}^{m-1}$, leads to the contour of the amoeba. In other words,
\begin{align}
 \Phi\brac{\mathbb{RP}^{m-1}}=\mathcal{C}\subset\mathcal{A}.
\end{align}
The diagram is then restricted to the following:
\begin{align*}
\xymatrix{      \mathbb{RP}^{m-1} \ar[dr]_{\Phi} \ar[r]^{\Psi}& W \ar[d]^{\mathrm{Log}} \\
      & \mathcal{C}
}
\end{align*}
Here, $W\subset D_B$ is the set of the critical points of the logarithmic map restricted to $D_B$, and $\mathcal{C}\subset\mathcal{A}$ is the contour of the amoeba. 

This latter situation is the one of relevance to circular geodesics, as the real-valued coordinates of $\mathbb{RP}^{m-1}$ will eventually be the (functions of the) coordinates of the light ring (for instance, the radius). Indeed, for polynomials of real variables, their discriminants are always the image of $\Phi(\mathbb{RP}^{m-1})=D_B$, and when we apply the methods to geodesics, we take $\Psi(\mathbb{RP}^{m-1})=\mathcal{C}$.

\subsection{The case \texorpdfstring{$m=2$}{m=2} and its consequences for circular geodesics} \label{subsec_Rusek}

In Refs.~\cite{Rusekthesis,Rusek2016}, Rojas and Rusek provided some specific results for the case $m=2$. In this case, the Horn--Kapranov map composed with the logarithm is $\Phi:\mathbb{CP}^1\rightarrow\mathbb{R}^2$, and the restriction of the map to $\mathbb{RP}^1$ gives a contour $\mathcal{C}$ of the amoeba of the $A$-discriminant, which is a one-dimensional curve in $\mathbb{R}^2$. Explicitly, the map is written as 
\begin{align}
 \Phi\brac{[\lambda_1:\lambda_2]}=\brac{\sum_{j=1}^{n+3}b_j^{1}\log\left|\langle b_j,\lambda\rangle \right|,\sum_{j=1}^{n+3}b_j^{2}\log\left|\langle b_j,\lambda\rangle \right|},
\end{align}
and $\langle b_j,\lambda\rangle=b_j^{1}\lambda_1+b_j^{2}\lambda_2$. The points where $\mathcal{C}$ are not differentiable appear as nodes and cusps. In \cite{Rusekthesis,Rusek2016}, Rojas and Rusek provided a theorem which gives an upper bound of the number of cusps in this case, which is given as follows.

\begin{mythm}[Rojas and Rusek \cite{Rusekthesis,Rusek2016}]
 In the case $m=2$ the graph of $\Phi$ has at most $n$ cusps.
\end{mythm}
For the univariate case $n=1$, this result states that the discriminant of a univariate polynomial which is a sum of $n+m+1=1+2+1=4$ monomials will have at most a single cusp. 

To see how this relates to circular geodesics, we consider spacetimes which can be cast in a form where its metric components depend only on a single variable, say, $x$. Then, the function $F(x)$ described in Eq.~\Eqref{prime_eqn2} can be cast in the form of an univariate polynomial if the metric components are rational functions of $x$. If, suppose, that $F(x)$ consists of a sum of four monomials, it takes the form 
\begin{align}
 F(x)&=ax^m+bx^n+x^k+x^l. \label{F_4mono}
\end{align}
Here, two of the coefficients can be set to $1$ by an appropriate choice of $h(x)$ and a rescaling of $x$. Circular orbits then correspond to $F(x)=F'(x)=0$, which gives
\begin{align}
 a=\frac{(n-k)x^k+(n-l)x^l}{x^m(m-n)},\quad b=\frac{-(m-k)x^k-(m-l)x^l}{x^n(m-n)}. \label{Fab}
\end{align}
This describes the $A$-discriminantal variety of a real polynomial as a curve in the $(a,b)$-plane, parametrised by $x$.  In the language of the Horn--Kapranov uniformisation, this is the image of $\mathbb{RP}^2$ under the map $\Phi$. Hence Eq.~\Eqref{Fab} describes $D_B$. We assume $h(x)$ to be positive. (The negative $h(x)$ case is the same upon reversing the signs and inequalities.) Consequently, along this curve, the circular orbits are stable if $F(x)$ is at a local maximum, i.e., $F''(x)<0$ and unstable if $F''(x)>0$. To identify the critical point, the additional condition $F''(x)=0$ leads to the simultaneous equations
\begin{subequations}\label{Fpp}
\begin{align}
 (mn-mk-kn+k^2)x^k+(mn-ml-ln+l^2)x^l&=0,\\
 (nk-nm-k^2+mk)x^k+(nl-nm-l^2+ml)x^l&=0.
\end{align}
\end{subequations}
On the other hand, tangents to the discriminant curve are obtained by taking the derivative of \Eqref{Fab} with respect to its parameter $x$, giving 
\begin{subequations}\label{apbp}
 \begin{align}
  a'&=-\frac{(nk+mk-k^2-mn)x^k+(nl+ml-l^2-mn)x^l}{x^{m+1}(m-n)},\\
  b'&=-\frac{(nk+mk-nm-k^2)x^k+(nl+ml-mn-l^2)x^l}{x^{n+1}(m-n)},
 \end{align}
\end{subequations}
where primes denote derivatives with respect to $x$. We observe that the numerators of $a'$ and $b'$ coincide with the left-hand sides of Eq.~\Eqref{Fpp}. Therefore, at the critical points $F''(x)=0$, we have $a'=b'=0$, and the tangents are undefined. Hence, $F''(x)=0$ correspond to cusps of the $A$-discriminant in the $(a,b)$-plane. 

Passing through a cusp means $F''(x)$ changes sign, therefore, a cusp of the discriminant curve separates two physically distinct branches of circular orbits, one of which has $F''(x)>0$ and the other $F''(x)<0$. If $F(x)$ consists of a sum of four monomials as we have assumed, then Rojas and Rusek's theorem asserts that the contour $\mathcal{C}$ of the amoeba of the Horn--Kapranov parametrisation has at most $n=1$ cusp. However, since the map $\mathrm{Log}$ between $D_B$ and $\mathcal{C}$ is a diffeomorphism for $a\neq0$ and $b\neq0$, this means $D_B$ itself also has at most 1 cusp. On $D_B$, the cusp separates two branches of circular orbits of opposite signs of $F''(x)$. We conclude that \emph{spacetimes whose geodesic effective potential takes the form \Eqref{F_4mono} have at most two branches of circular orbits, one of which is stable and the other unstable}.

\section{Example: Time-like geodesics in Schwarzschild spacetime} \label{sec_Sch}

As an illustrative example demonstrating the concepts of the previous discussions, let us consider the well-known problem of time-like geodesics in the Schwarzschild spacetime, where the metric is given by 
\begin{subequations} \label{metric_Sch}
\begin{align}
 \dif s^2&=-f\dif t^2+\frac{\dif r^2}{f}+r^2\dif\theta^2+r^2\sin^2\theta\,\dif\phi^2,\\
     f&=1-\frac{2M}{r},
\end{align}
\end{subequations}
where $M>0$ is the mass parameter of the black hole. Here $\frac{\partial}{\partial t}$ and $\frac{\partial}{\partial\phi}$ are the Killing vectors with the corresponding conserved quantities, 
\begin{align}
 P_t=-E=-f\dot{t},\quad P_\phi=L=r^2\sin^2\theta\dot{\phi},
\end{align}
which we regard as the energy and angular momentum of the particle, respectively. 
Due to the spherical symmetry of the spacetime, we may take without loss of generality $\theta=\frac{\pi}{2}=\mbox{constant}$. We also focus on time-like geodesics, $\epsilon=-1$. Then, in the present case 
\begin{align}
 -\mathcal{U}&=\frac{1}{f}\sbrac{E^2-\frac{L^2}{r^2}f-f}\nonumber\\
    &=\frac{1}{1-2M/r}\brac{\frac{2M}{r}}^3\brac{\frac{L^2}{4M^2}}\sbrac{\frac{4M^2}{L^2}\brac{1-E^2}\brac{-\frac{r}{2M}}^3+\frac{4M^2}{L^2}\brac{\frac{r}{2M}}^2-\frac{r}{2M}+1}.
\end{align}
Defining $x=-\frac{r}{2M}$, the equation in the square brackets above is the degree-3 polynomial
\begin{align}
 F(x)&=ax^3+bx^2+x+1, \label{Sch_F}
\end{align}
where 
\begin{align}
 a=\frac{4M^2(1-E^2)}{L^2},\quad b=\frac{4M^2}{L^2}.
\end{align}
This $F(x)$ is a univariate polynomial which is a sum of four monomials, and hence fits the conditions of Rojas and Rusek's theorem.

Inverting the above equation in terms of the conserved quantities, we have 
\begin{align}
 E^2=\frac{b-a}{b},\quad \frac{L^2}{4M^2}=\frac{1}{b}.
\end{align}
Finding the discriminant using $F(x)=F'(x)=0$, we have 
\begin{align}
 a=\frac{2+x}{x^3},\quad b=-\frac{3+2x}{x^2}. \label{Sch_discr1}
\end{align}

On the other hand, we apply the Horn--Kapranov uniformisation by noting that the set of exponents of $F$ is $A=\{0,1,2,3\}$. Hence the corresponding $A$-matrix and its Gale dual are
\begin{align}
 A=\left(\begin{array}{cccc}
                1 & 1 & 1 & 1\\
                0 & 1 & 2 & 3                
               \end{array}\right),\quad 
    B=\left(\begin{array}{rr}
          2 & 1 \\ -3 & -2 \\ 0 & 1 \\ 1 & 0
         \end{array}\right).
\end{align}
The Horn--Kapranov map \Eqref{psi_def} gives the reduced $A$-discriminant as 
\begin{align}
 \Psi\brac{[\lambda_1:\lambda_2]}=\brac{\frac{\lambda_1(2\lambda_1+\lambda_2)^2}{(-3\lambda_1-2\lambda_2)^3},\,\frac{\lambda_2(2\lambda_1+\lambda_2)}{(-3\lambda_1-2\lambda_2)^2}}.
\end{align}
Taking a patch of $\mathbb{RP}^1$ where $\lambda_1\neq 0$, we have $[\lambda_1:\lambda_2]=[1:\lambda]$, and we have $\Psi\brac{[1:\lambda]}=(a,b)$, where
\begin{align}
 a&=\frac{(2+\lambda)^2}{(-3-2\lambda)^3},\quad b=\frac{\lambda(2+\lambda)}{(-3-2\lambda)^2}. \label{Sch_discr2}
\end{align}
Eliminating $\lambda$ between $a$ and $b$ leads to $b^2-4a-4b^3-27a^2+18ab=0$, which is precisely the vanishing discriminant condition for \Eqref{Sch_F}. Composing this with the log map gives the amoeba contour $\mathcal{C}$,
\begin{align}
 \Phi\brac{[1:\lambda]}=\brac{\log|a|,\,\log|b|},
\end{align}
where $a$ and $b$ are as defined in Eq.~\Eqref{Sch_discr2}. 

Comparing Eqs.~\Eqref{Sch_discr1} and \Eqref{Sch_discr2}, we find $x=-\frac{3+2\lambda}{2+\lambda}$, or recalling $x=-\frac{r}{2M}$,
\begin{align}
 \lambda=\frac{3-\frac{r}{M}}{\frac{r}{2M}-2} \quad\leftrightarrow\quad\frac{r}{2M}=\frac{3+2\lambda}{2+\lambda}.
\end{align}
This relates the parameter $\lambda$ to the radius $r$ of the circular geodesics. We note that the cusp of $D_A$ occurs at $\lambda=-3$, which corresponds to $r=6M$, hence recovering the \emph{innermost stable circular orbit} (ISCO) around the Schwarzschild black hole. This is the point where stable circular orbits $r>6M$ turn into unstable ones at $r<6M$.

\begin{figure}
 \centering
 \begin{subfigure}[b]{0.49\textwidth}
    \centering
    \includegraphics[width=0.6\textwidth]{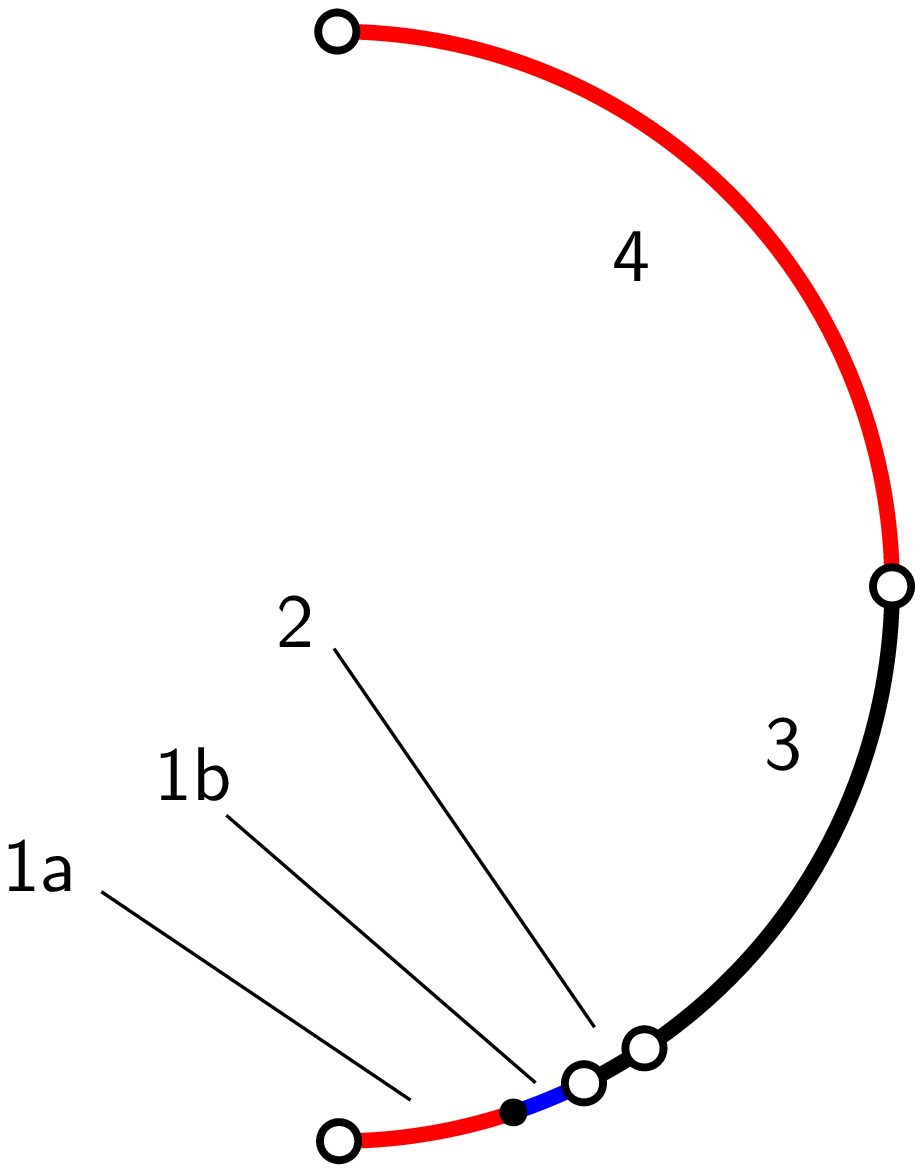}
    \caption{Parametrising $\mathbb{RP}^1$ in $S^1_+$.}
    \label{fig_Sch_d}
  \end{subfigure}
 \begin{subfigure}[b]{0.49\textwidth}
    \centering
    \includegraphics[width=\textwidth]{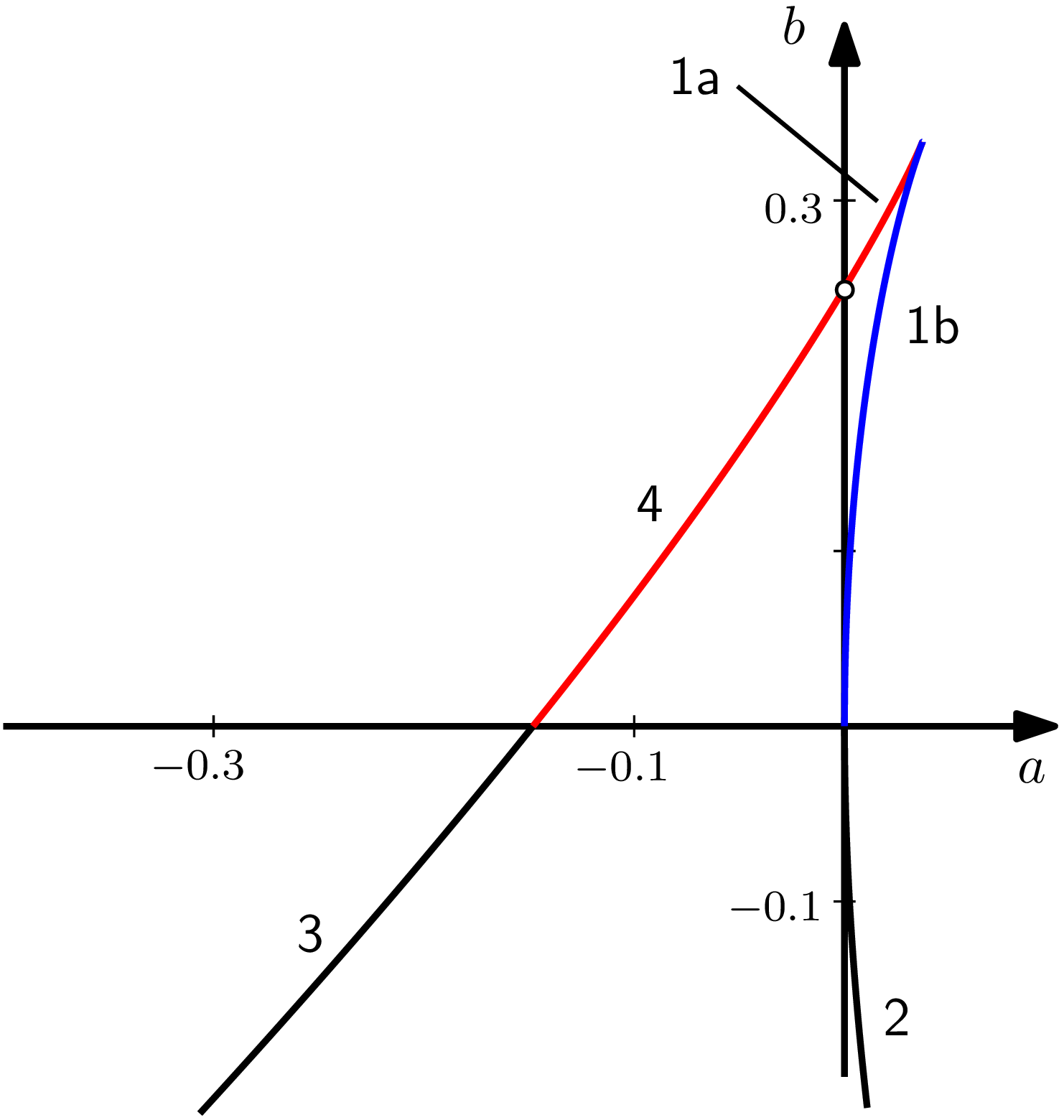}
    \caption{Reduced $A$-discriminant $D_B$.}
    \label{fig_Sch_a}
  \end{subfigure}
  \par 
  \vspace{12pt}
  \begin{subfigure}[b]{0.49\textwidth}
    \centering
    \includegraphics[width=\textwidth]{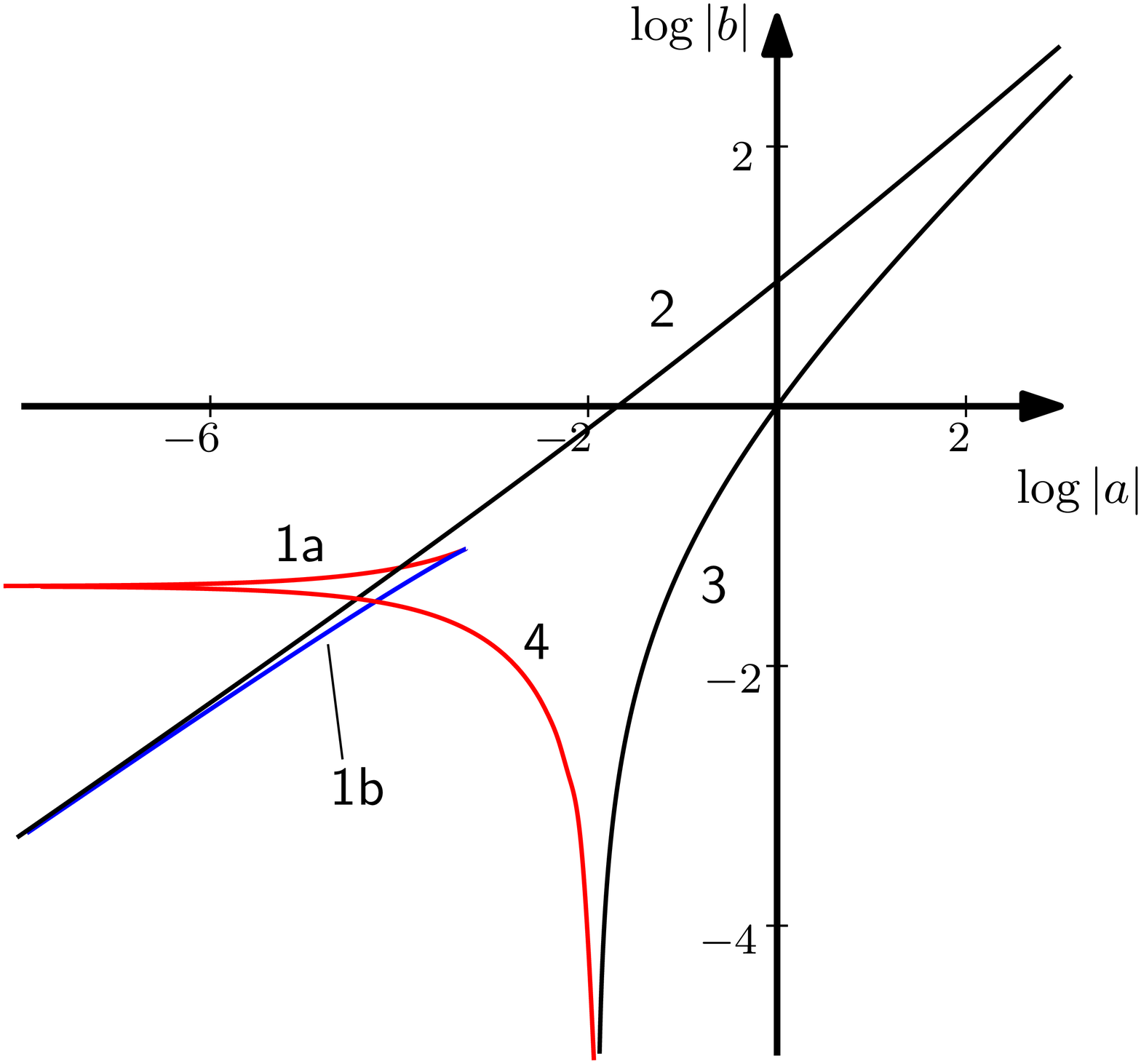}
    \caption{Contour, $\mathcal{C}$.}
    \label{fig_Sch_b}
  \end{subfigure}
  \begin{subfigure}[b]{0.49\textwidth}
    \centering
    \includegraphics[width=\textwidth]{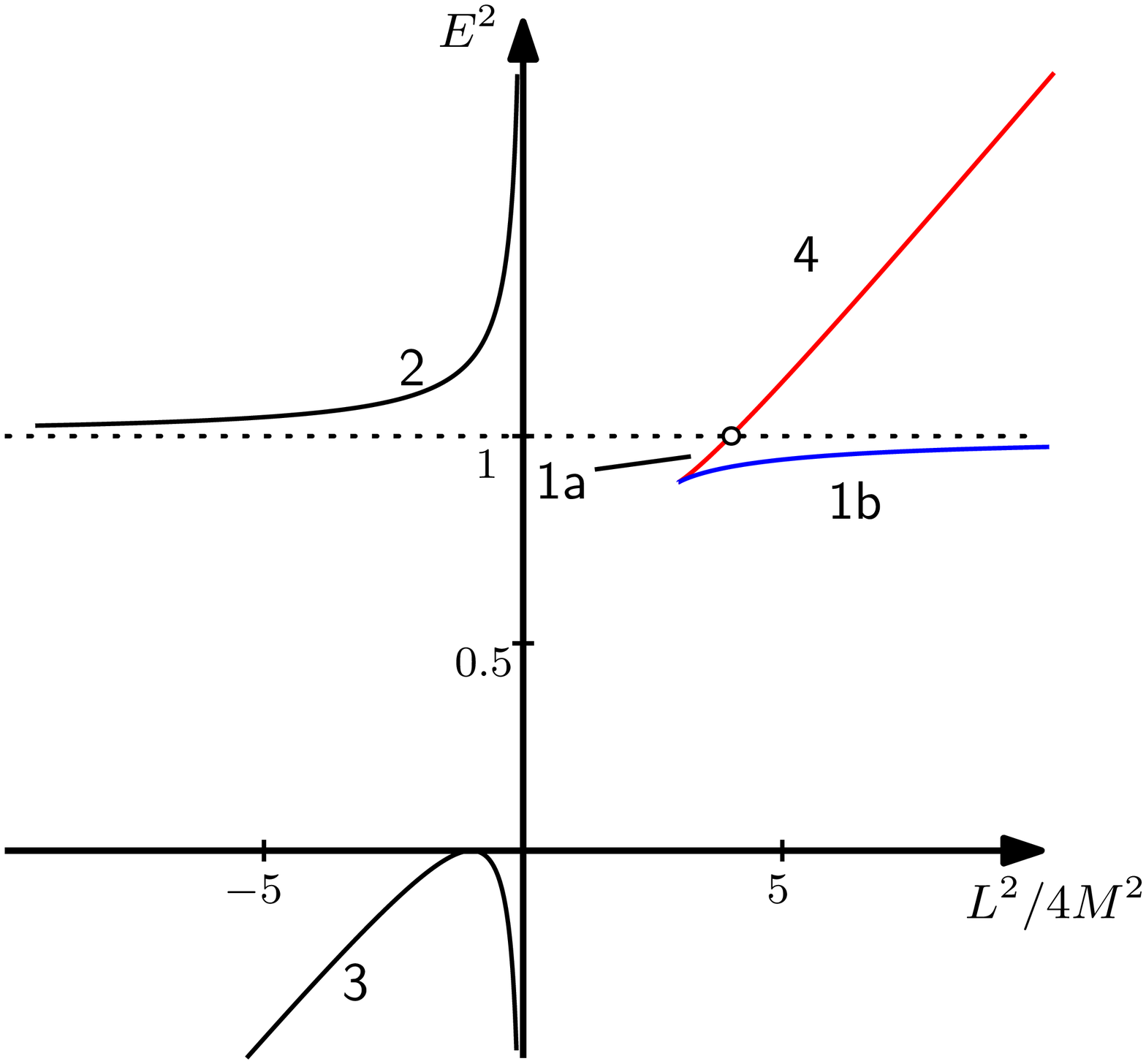}
    \caption{$E^2$ vs $L^2/4M^2$.}
    \label{fig_Sch_c}
  \end{subfigure}
  \caption{The semicircle parametrisation of $\mathbb{RP}^1$ (\ref{fig_Sch_d}) and the reduced $A$-discriminant (\ref{fig_Sch_a}) and contour (\ref{fig_Sch_b}) of the corresponding amoeba in the problem of time-like circular orbits in the Schwarzschild spacetime. The plot in terms of physical parameters $E^2$ and $L^2/4M^2$ is included (\ref{fig_Sch_c}). The semicircle is subdivided into domains \textsf{1a}--\textsf{4}, and their corresponding images are marked, respectively, in the other three graphs. The blue curves (Domain \textsf{1b}) represent the stable branch of circular orbits and the red curves (Domain \textsf{1a} and \textsf{4}) indicate the unstable branch. The black curves (Domains \textsf{2} and \textsf{3}) indicate unphysical circular orbits.}
  \label{fig_Sch}
\end{figure}

This map is undefined for $\lambda=-\frac{3}{2}$, $\lambda=-2$, and $\lambda=0$. Therefore we divide the possible values of $\lambda$ into open domains separated by these points. Furthermore, we subdivide the domain containing the ISCO ($\lambda=-3$) into the stable and unstable parts:
\begin{align}
 \mbox{\textsf{Domain 1a}:}\quad & -\infty<\lambda <-3, \quad\mbox{(unstable circular orbits)}, \nonumber\\
 \mbox{\textsf{Domain 1b}:} \quad & -3<\lambda<-2, \quad\mbox{(stable circular orbits)},\nonumber\\
 \mbox{\textsf{Domain 2}:}\quad & -2<\lambda<-\frac{3}{2}, \quad\mbox{(unphysical)}, \nonumber\\
 \mbox{\textsf{Domain 3}:}\quad & -\frac{3}{2}<\lambda<0, \quad\mbox{(unphysical)},\nonumber\\
 \mbox{\textsf{Domain 4}:}\quad & 0<\lambda<\infty, \quad\mbox{(unstable circular orbits)}.
\end{align}
We can visualise $\mathbb{RP}^1$ by parametrising it using a semi-circle, the domains are as depicted in Fig.~\ref{fig_Sch_d}, where the angles along the semi-circle $S^1_+$ are given by $\theta=\arctan\lambda$.

In terms of $r$, Domain \textsf{1a} corresponds to $4M<r<6M$, Domain \textsf{1b} corresponds to $6M<r<\infty$, and Domain \textsf{4} corresponds to $3M<r<4M$. Domains \textsf{2} and \textsf{3} are unphysical as they correspond to negative $r$.

The reduced $A$-discriminant $D_B$ and contour $\mathcal{C}$ are depicted in Figs.~\ref{fig_Sch_a} and \ref{fig_Sch_b}, respectively, where we can clearly see the presence of a single cusp at $\lambda=-3$ or $r=6M$. The five segments labelled \textsf{1a}--\textsf{4} are the corresponding images of Domains \textsf{1a}--\textsf{4} under $\Psi$ and $\Phi$, respectively. In each figure, the unphysical Domains \textsf{2} and \textsf{3} along with their corresponding images are drawn in black, whereas Domain \textsf{1b} and its images are drawn in blue, representing stable circular orbits. Domains \textsf{1a} and \textsf{4} are depicted in red, representing unstable circular orbits. Domains \textsf{1a} and \textsf{1b} meet at a cusp corresponding to $r=6M$, the ISCO. 

Finally, we note that $\lambda\rightarrow\pm\infty$ corresponds to $r\rightarrow 4M$. This is the circular orbit of energy $E=1$. This case is undefined in the above map as it corresponds to $a=0$, and the leading term of the polynomial vanishes. Physically, this is the critical point between the bound energy $(E<1)$ and the unbound ones $(E>1)$. Unstable circular orbits of $3M<r<4M$ carries energy $E>1$ and may escape to infinity upon perturbation.

\section{Asymptotically flat spacetimes with at most two light rings} \label{sec_classes}

We now turn our attention to null geodesics in static, spherically symmetric spacetimes. Our task is to identify a class of spacetimes for which the equations for null geodesics satisfy the conditions of Rojas and Rusek's theorem. For static, spherically symmetric spacetimes, the effective potential will ultimately be a function of a single radial variable. It then corresponds to $n=1$ in the context of Sec.~\ref{subsec_Rusek}, if our equations are algebraic. Then Rojas and Rusek's theorem implies that such spacetimes can have up to two light rings, one of which is stable and the other unstable.

\subsection{Setup} \label{subsec_general}

Let us consider static, $D$-dimensional spherically symmetric spacetimes, where the metric can be written in the form 
\begin{align}
 \dif s^2&=-f(r)\dif t^2+h(r)\dif r^2+r^2\gamma_{ij}\dif\phi^i\dif\phi^j,
\end{align}
where $f(r)$ and $h(r)$ are functions of $r$ only and $\gamma_{ij}\dif\phi^i\dif\phi^j=\dif\Omega^2_{(D-2)}$ is the metric of $S^{D-2}$, the unit $(D-2)$-sphere. The spacetime is asymptotically flat if $\lim_{r\rightarrow\infty} f(r)=\lim_{r\rightarrow\infty} h(r)=1$ and describes a black hole if there exists a horizon at $r=r_{\mathrm{H}}$ where $f(r_{\mathrm{H}})=0$. The horizon is extremal if this root is degenerate. On the other hand, the spacetime may describe a horizonless compact object if there is no root $f(r)=0$ for $r>0$.

The Lagrangian for geodesic motion is
\begin{align*}
 \mathcal{L}&=\half\brac{-f\dot{t}^2+h\dot{r}^2+r^2\gamma_{ij}\dot{\phi}^i\dot{\phi}^j}.
\end{align*}
Performing the Legendre transform to obtain the Hamiltonian, its corresponding Hamilton--Jacobi equation is 
\begin{align}
 \half\sbrac{-\frac{1}{f}\brac{\frac{\partial S}{\partial t}}^2+\frac{1}{h}\brac{\frac{\partial S}{\partial r}}^2+\frac{1}{r^2}\gamma^{ij}\frac{\partial S}{\partial\phi^i}\frac{\partial S}{\partial\phi^j}}+\frac{\partial S}{\partial\tau}=0.
\end{align}
We take our ansatz to be $S=-\half\epsilon\tau-Et+R(r)+\Phi\brac{\phi^1,\ldots,\phi^{D-2}}$. Then, we find there exists a separation constant $L$ such that the equation for $R$ is
\begin{align}
 \frac{1}{h}\brac{\frac{\dif R}{\dif r}}^2=\frac{p_r^2}{h}=\frac{1}{f}\brac{E^2-\frac{L^2}{f}+\epsilon f}.
\end{align}
The remaining equations are $\gamma^{ij}\frac{\partial\Phi}{\partial\phi^i}\frac{\partial\Phi}{\partial\phi^j}=L^2$ which describe the angular motions, which we will not consider here in detail as they depend mainly on the spherical geometry of the spacetime, rather than its gravitating source. However, we do point out that in the case $D=4$, it was shown in Ref.~\cite{Cunha:2017qtt} that light rings are stable along the angular directions. 

For null geodesics, we have $\epsilon=0$, and the effective potential for radial motion is
\begin{align}
 -\mathcal{U}=\frac{L^2}{f}\brac{\eta-\frac{1}{r^2}f},\quad \eta=\frac{E^2}{L^2}. \label{genF1}
\end{align}
We assume $f$ to be a rational function of the form
\begin{align}
 f=1-\frac{P(r)}{Q(r)}, \label{fPQ}
\end{align}
where $P(r)$ and $Q(r)$ are polynomials in $r$. For asymptotically flat spacetimes, we require $\lim_{r\rightarrow\infty}\frac{P}{Q}=0$. For this form of $f$, Eq.~\Eqref{genF1} now takes the form $-\mathcal{U}=\frac{L^2}{r^2Qf}\mathcal{F}(r)$, where
\begin{align}
 \mathcal{F}(r)&=\eta r^2Q-Q+P.
\end{align}
We assume that the prefactor $\frac{L^2}{r^2Qf}$ will always be positive in the domain of geodesic motion. A light ring then corresponds to $\mathcal{F}(r)=\mathcal{F}'(r)=0$, which defines the $A$-discriminantal variety of $\mathcal{F}$.

Of course, applying Einstein's equation or any alternative theory of gravity's field equation would determine $f$ that may or may not take the form \Eqref{fPQ}. However, we are not presently invoking any particular model of gravity other than the above-mentioned general assumptions of our spacetime. Instead, we are considering classes of functions $f$ that fit the conditions so that Rojas and Rusek's theorem can be applied. In particular, $\mathcal{F}(r)$ must be a sum of four monomials. To achieve this, we have a few cases. In the following we take the exponents $m$, $n$, and $l$ to be distinct, non-negative integers, and $A$, $B$, and $C$ to be real coefficients.
\begin{enumerate} 
 \item \textbf{Class I.} If $Q$ is a sum of two distinct monomials whose degrees differ by an integer greater than 2, then $\eta^2 r^2Q-Q$ already consists of four distinct monomials. Hence $P$ must have monomials in the same degree as those in $r^2Q$ or $Q$. Requiring asymptotic flatness immediately rules out the highest degree terms in both. Therefore, $f$ takes the form 
 \begin{align}
  f_{\mathrm{I}}=1-\frac{B r^{n+2}+C r^n}{r^m+A r^n}, \quad m>n+2. \label{ClassI}
 \end{align}

 \item \textbf{Class II.} In this case, $Q$ is a sum of two distinct monomials whose degrees differ by 2. For asymptotic flatness, $P$ cannot have any of the same monomials in $r^2Q$, nor the leading monomial of $Q$. Therefore, $f$ takes the form
 \begin{align}
  f_{\mathrm{II}}=1-\frac{Br^l}{r^m+Ar^{m-2}},\quad m>l,\quad l\neq m-2. \label{ClassII}
 \end{align}
 
 \item \textbf{Class III.} Here, $Q$ is a sum of two distinct monomials whose degrees differ by 1, 
 \begin{align}
  f_{\mathrm{III}}=1-\frac{Br^{l}}{r^m+Ar^{m-1}},\quad m-1>l. \label{ClassIII}
 \end{align}

 \item \textbf{Class IV.} $Q$ has only 1 term. Then, $P$ has two distinct monomials whose degrees are less than $Q$ to ensure asymptotic flatness.
 \begin{align}
  f_{\mathrm{IV}}=1-\frac{Ar^n+Br^l}{r^m},\quad n,l<m. \label{ClassIV}
 \end{align}
\end{enumerate}

\subsection{Absence of stable light rings in spacetimes with non-degenerate horizons}

We now show that in the four classes $f$ identified above, the parameters supporting the presence of a stable light ring correspond to spacetimes with no non-degenerate horizon. In Refs.~\cite{Khoo:2016xqv,Tang:2017enb}, it was shown that certain spacetimes with a degenerate horizon carry a stable light ring on the location of the horizon.

This latter statement can be shown in a direct manner when we restrict our attention to the  Classes I--IV functions defined in Sec.~\ref{subsec_general}. First, observe that in the limit $\eta\rightarrow0$, the $A$-discriminantal variety of $\mathcal{F}$ coincides with that of $-(Q-P)$, which is where $f=1-\frac{P}{Q}$ has a degenerate root. Therefore the zero-energy limit of a light ring coincides with the extremal horizon of the black hole. In this limit, $\mathcal{F}''(r)=-(Q''(r)-P''(r))\propto -f''(r)$. If this were to be the extremal horizon of an asymptotically flat spacetime, then $f''(r)<0$ at the horizon. Therefore $\mathcal{F}''(r)>0$ there. This means that extremal black holes carry stable light rings on their horizons, albeit for photons with zero energy.

Next, we show that if we follow the $A$-discriminantal variety of $\mathcal{F}$ continuously as $\eta>0$, the parameters take values such that the degenerate roots of $f$ are taken into the complex domain. Therefore, the horizon disappears where a stable light ring exists. This can be demonstrated for each class explicitly.

\textbf{Class I.} We begin with Class I solutions, where 
\begin{align}
 f=1-\frac{P}{Q},\quad P=Br^{n+2}+Cr^n,\quad Q=r^m+Ar^n.
\end{align}
In order to have attractive gravity as $r$ increases towards infinity, we require $B>0$. Also, to avoid curvature singularities beyond $r>0$, we have $A>0$. The horizons corresponds to the roots
\begin{align}
 f=0\quad\leftrightarrow\quad Q-P=r^m-Br^{n+2}+(A-C)r^n=0. \label{QPI}
\end{align}
A degenerate root will occur for 
\begin{align}
 \frac{B^{m-n}}{(A-C)^{m-n-2}}=\zeta,\quad\mbox{where }\quad \zeta=\frac{(m-n)^{m-n}}{4(m-n-2)^{m-n-2}}. \label{ClassIzeta}
\end{align}
The degenerate root becomes complex when $\frac{B^{m-n}}{(A-C)^{m-n-2}}<\zeta$, and becomes real when $\frac{B^{m-n}}{(A-C)^{m-n-2}}>\zeta$.

We now turn to the function $\mathcal{F}(r)$ characterising the geodesic motion. As before, light rings correspond to the discriminantal variety of $\mathcal{F}(r)=\eta r^2Q-Q+P$ determined by $\mathcal{F}(r)=\mathcal{F}'(r)=0$. For the values of $\eta$ and $B$ satisfying this condition, we find that its second derivative is
\begin{align}
 \left.\mathcal{F}''(r)\right|_{\mathrm{lightring}}&=-\frac{2}{r^2}\sbrac{(m-n+2)(A-C)r^n-(m-2-n)r^m}.
\end{align}
If $A<C$, the terms inside the square bracket are always negative; hence, $\mathcal{F}''(r)$ is always positive. Therefore, the light ring in this case is always unstable. 

On the other hand, if $A>C$, the light ring may or may not be unstable. We show that the stable case always corresponds to spacetimes with no horizons. We let $r=(A-C)^{\frac{1}{m-n}}x$, then $\mathcal{F}$ can be written as 
\begin{align}
 \mathcal{F}=(A-C)^{\frac{m}{m-n}}F(x),
\end{align}
where
\begin{align}
 F(x)&=ax^{m+2}+bx^{n+2}-x^m-x^n,\nonumber\\
  a&=\eta (A-C)^{\frac{2}{m-n}},\quad b=\eta A(A-C)^{-\frac{m-n-2}{m-n}}+B(A-C)^{-\frac{m-n-2}{m-n}}. \label{ClassIF}
\end{align}
The discriminantal variety of $F$ is given by the solution of $F(x)=F'(x)=0$:
\begin{align}
 a=\frac{(m-n-2)x^m-2x^n}{x^{m+2}(m-2)},\quad b=\frac{2x^m+(m-n+2)x^n}{x^{n+2}(m-n)}. \label{ClassIab}
\end{align}
We note that the tangent slope of this discriminantal variety is
\begin{align}
 \frac{\dif b}{\dif a}=\frac{b'}{a'}=-x^{m-n}, \label{ClassIdbda}
\end{align}
which is always negative. In particular, we consider the limit $\eta\rightarrow0$, which corresponds to the point $a=0$ and where $Q-P$ and $\mathcal{F}$ share the same degenerate root. Hence at this point, $b=\frac{B}{(A-C)^{\frac{m-n-2}{m-n}}}=\zeta^{1/(m-n)}$.

We now follow the curve continuously along Eq.~\Eqref{ClassIab}, as $a$ increases positively from zero, where $\mathcal{F}$ continues to have the degenerate root (describing the stable light rings). However, Eq.~\Eqref{ClassIdbda} means that $b$ must decrease. Looking at the equation for $b$ in Eq.~\Eqref{ClassIF}, this means $B(A-C)^{-\frac{m-n-2}{m-n}}$ must decrease below its value when $\eta=0$, which was $\zeta^{1/(m-n)}$. This means entering $a>0$ leads to the degenerate roots of $f=(Q-P)/Q$ becoming complex, and hence, the horizon vanishes along the branch of stable light rings. By the Descartes rule of signs, Eq.~\Eqref{QPI} can have at most two positive real roots. So when they are the degenerate root/horizon which vanishes as we increase $\eta$, we are left with light rings in a horizonless spacetime. Now, $b$ will continue to decrease along this branch until $b'=0$, after which $b$ may increase again. But this is the point $F''(x)=0$. So after this point, we then encounter the unstable branch of light rings.

\textbf{Class II.} In this class, we have 
\begin{align}
 f_{\mathrm{II}}=1-\frac{P}{Q},\quad P=Br^l,\quad Q=r^m+Ar^{m-2}.
\end{align}
As in Class I, to have a weakening attractive gravity as $r$ grows, along with no curvature singularities at $r>0$ requires $B$ and $A$ to be positive. 
Looking at the discriminant condition $\mathcal{F}(r)=\mathcal{F}'(r)=0$ for $\mathcal{F}=\eta r^2Q-Q+P$, we find that for $\eta$ and $B$ obeying this condition leads to the second derivative 
\begin{align}
 \left.\mathcal{F}''(r)\right|_{\mathrm{lightring}}=2(m-2-l)\left\{ \frac{(m-l)A^2 r^{m-2}+2(m+2-l)Ar^m+(m-l)r^{2m}}{r^2\sbrac{(m-l)A+(m+2-l)r^2}}\right\}.
\end{align}
If $l<m-2$, the terms inside the curly braces are always positive, along with the prefactor. Therefore, the second derivative is always positive and the corresponding light rings are unstable. 

The case $m<l<m-2$ just means $l=m-1$. In this case, we find
\begin{align}
 Q-P=r^{m-2}\brac{r^2-Br+A};
\end{align}
therefore, a degenerate root occurs when $B^2=4A$. This degenerate root becomes complex as $B^2<4A$. 

Letting $r=\frac{A}{B}x$, the function $\mathcal{F}$ can be cast in the form 
\begin{align}
 \mathcal{F}=\frac{A^{m-1}}{B^{m-2}}x^{m-2}F(x),
\end{align}
where $F(x)=ax^4+bx^2+x-1$,
\begin{align}
 F(x)=ax^4+bx^2+x-1,\quad a=\eta\frac{A^3}{B^4},\quad b=(\eta A-1)\frac{A}{B^2}. \label{ClassIIab}
\end{align}
The discriminantal variety of $\mathcal{F}$ is then equivalent to the discriminantal variety of $F(x)$, which is now given by 
\begin{align}
 a=\frac{x-2}{2x^4},\quad b=\frac{4-3x}{2x^2}.
\end{align}
As in the general argument at the beginning of this section, the degenerate root of $F$ coincides with that of $Q-P$ in the limit $\eta\rightarrow 0$, and this corresponds to a stable light ring at $a=0$, in which case $b=-\frac{A}{B^2}=-\frac{1}{4}$. Increasing $\eta$ continuously from zero into positive values means following the discriminantal variety continuously as $a>0$. However, we find that 
\begin{align}
 \frac{b'}{a'}=\frac{\dif b}{\dif a}=-x^2.
\end{align}
This means $b$ must decrease as we follow this curve. Looking at the expression for $b$ in Eq.~\Eqref{ClassIIab}, the prefactor $(\eta A-1)$ is a negative term decreasing in magnitude as we increase $\eta$ from zero. This means $\frac{A}{B^2}$ must increase from its previous value at $a=0$ (namely $\frac{A}{B^2}=\frac{1}{4}$.) which takes the degenerate root of $Q-P$ into the complex domain. Therefore, the horizon vanishes along the branch of stable light rings.

\begin{figure}
 \centering 
 \includegraphics[width=0.49\textwidth]{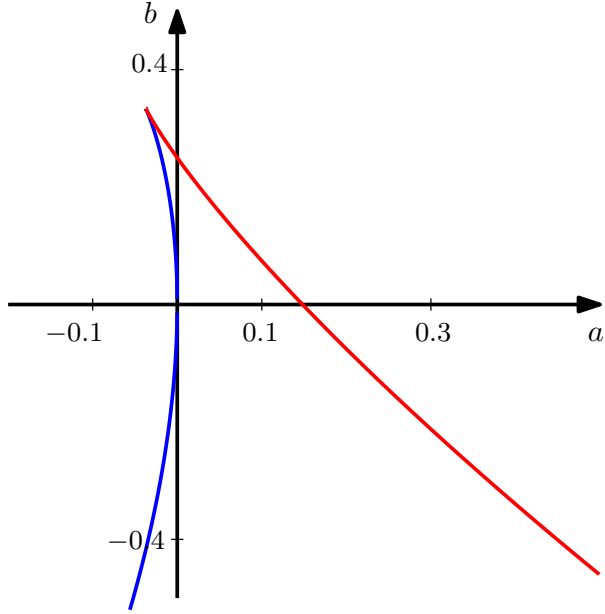}
 \caption{Discriminant curve for the $F$-function due to the Class III function.}
 \label{fig_Class2_ab}
\end{figure}

\textbf{Class III.} In this case, we have $P=Br^{m-1}$ and $Q=r^m+Ar^{m-1}$. Horizons occur if 
\begin{align}
 Q-P=r^{m-1}\sbrac{r-(B-A)}=0.
\end{align}
Therefore, the spacetime has a horizon if $B>A$ and no horizon if $B<A$. Letting $r=(B-A)x$, the function $\mathcal{F}$ takes the form 
\begin{align}
 \mathcal{F}=(B-A)^mF(x),
\end{align}
where 
\begin{align}
 F(x)=ax^3+bx^2-x+1,\quad a=\eta(B-A)^2,\quad b=\eta A(B-A).
\end{align}
The discriminantal variety is given by 
\begin{align}
 a=\frac{2-x}{x^3},\quad b=\frac{2x-3}{x^2}.
\end{align}
The curve is depicted in Fig.~\ref{fig_Class2_ab}, where we find that the stable branch (blue) is in the unphysical $a<0$ region corresponding to $\eta<0$. The physically feasible region then is in the $a>0$ part, with $b>0$ corresponding to parameters with a horizon and $b<0$ a spacetime with no horizon. In either case, we have the red branch representing unstable circular orbits. 

\textbf{Class IV.} In this case, $P=Ar^n+Br^l$ and $Q=r^m$. In order to have a weakening attractive gravity as $r\rightarrow\infty$, we assume $n>l$ and take $A>0$.  The horizons occur if 
\begin{align}
 Q-P=r^m-Ar^n-Br^l=0. \label{QPIV}
\end{align}
We will have degenerate roots for 
\begin{align}
 \frac{B^{m-n}}{A^{m-l}}=\zeta,\quad\mbox{ where }\quad \zeta=(-1)^{m-n}\frac{(m-n)^{m-n}(n-l)^{n-l}}{(m-l)^{m-l}}.
\end{align}
The degenerate root becomes complex as $\frac{B^{m-n}}{A^{m-l}}<\zeta$ and becomes real as $\frac{B^{m-n}}{A^{m-l}}>\zeta$.

Turning now to $\mathcal{F}(r)=\eta r^2Q-Q+P$ which characterises the geodesic motion, one can use the light-ring condition $\mathcal{F}(r)=\mathcal{F}(r)=0$ to determine the discriminantal variety. For the values of $\eta$ and $A$ obeying this condition, we find the second derivative to be 
\begin{align}
 \left.\mathcal{F}''(r)\right|_{\mathrm{lightring}}=\frac{2(m-n)r^m+(n-l)(m-l-2)Br^l}{r^2}.
\end{align}
Hence, the light ring is always unstable if $B>0$. However it is possibly unstable in the case $B<0$. Let us consider this latter case in further detail. Let $K=-B$ be a positive quantity. The roots of $Q-P$ are degenerate when 
\begin{align}
 \frac{K^{m-n}}{A^{m-l}}&=\zeta,\quad\mbox{ where }\quad \zeta=\frac{(m-n)^{m-n}(n-l)^{n-l}}{(m-l)^{m-l}}.
\end{align}
The degenerate root becomes complex as $\frac{K^{m-n}}{A^{m-l}}<\zeta$ and becomes real and distinct as $\frac{K^{m-n}}{A^{m-l}}>\zeta$.

Turning to the function $\mathcal{F}=\eta r^2Q-Q+P$, we let $r=A^{1/(m-n)}x$ so that it can be written as 
\begin{align}
 \mathcal{F}&=A^{\frac{m}{m-n}}F(x),
\end{align}
where 
\begin{align}
 F(x)=ax^{m+2}-x^m+x^n-bx^l,\quad a=\eta A^{\frac{2}{m-n}},\quad b=\frac{K}{A^{\frac{m-l}{m-n}}}.
\end{align}
We apply the same argument as in the previous classes. At $\eta=0$, $\mathcal{F}$ is a multiple of $Q-P$; hence, they both have a degenerate root at $b=K/A^{\frac{m-l}{m-n}}=\zeta$. We now follow the discriminantal variety of $F(x)=F'(x)=0$, which is given by 
\begin{align}
 a=\frac{(m-l)x^m-(n-l)x^n}{(m-l+2)x^{m+2}},\quad b=\frac{-2x^m+(m-n+2)x^n}{(m-l+2)x^l}.
\end{align}
The tangent to the curve in the $(a,b)$-plane is 
\begin{align}
 \frac{b'}{a'}=\frac{\dif b}{\dif a}=x^{m-l+2}, \label{ClassIVdbda}
\end{align}
which is always positive for $x>0$. Now suppose we are at $\eta=0$; then, the light ring occurs when $a=0$ and coincides with the extremal horizon, for which case $b=K/A^{\frac{m-l}{m-n}}=\zeta$. We now follow the discriminantal variety as $a$ increases continuously from zero to positive values. From Eq.~\Eqref{ClassIVdbda}, $b=K/A^{\frac{m-l}{m-n}}$ increases as well. Then, the previously degenerate roots of $Q-P$ become complex and the horizon vanishes. By the Descartes rule of signs for $A>0$ and $B<0$, Eq.~\Eqref{QPIV} can have at most two positive roots. When they become degenerate and vanish as $\eta$ increases, we are again left with stable light rings in a horizonless spacetime.

\section{Examples} \label{sec_examples}

We now turn to specific examples of known spacetimes which embody the consequences of Rojas and Rusek's theorem leading to two branches of light rings. In the following, we consider the Hayward and Reissner--Nordstr\"{o}m spacetime in some detail. We see how the general arguments of the previous section applies using the concrete parameters of the spacetimes in question.

\subsection{The Hayward spacetime} 

The Hayward spacetime \cite{Hayward:2005gi} is a four-dimensional spacetime with the metric given by 
\begin{subequations}\label{Hayward_metric}
\begin{align}
 \dif s^2&=-f\dif t^2+f^{-1}\dif r^2+r^2\dif\theta^2+r^2\sin^2\theta\,\dif\phi^2,\\
  f&=1-\frac{2Mr^2}{r^3+2M\ell^2}.
\end{align}
\end{subequations} 
Here, $M$ parametrises the mass of the central gravitating source, and the essential feature of this spacetime is that it lacks a curvature singularity at the origin for non-zero $\ell$.
The geodesics of the Hayward spacetime have been studied in \cite{Abbas:2014oua,Wei:2015qca,Chiba:2017nml,Hu:2018old}. A generalisation of the spacetime to include quintessence was recently studied in \cite{Pedraza:2020uuy}. In these works, it was found that for parameters of the spacetime containing a horizon (the Hayward black hole), there exists an unstable light ring. For parameters where the horizon vanishes, there are two branches of light rings, one stable and the other unstable \cite{Wei:2015qca,Chiba:2017nml}. In light of the discussion of $A$-discriminants in this paper, we  now reinterpret these results as a consequence of Rojas and Rusek's theorem. 

The Hayward metric \Eqref{Hayward_metric} is an example of a Class I spacetime with $A=2M\ell^2$, $B=2M$, $C=0$, $m=3$, and $n=0$. For these numbers, the critical value $\zeta$ as defined in Eq.~\Eqref{ClassIzeta} is $\zeta=\frac{27}{4}$. The condition for the degenerate roots becoming complex in terms of the present parameters as 
\begin{align}
 \frac{B^{m-n}}{(A-C)^{m-n-2}}<\zeta\quad\rightarrow\quad \frac{4M^2}{\ell^2}&<\frac{27}{4},
\end{align}
recovering the statements of the original analysis \cite{Hayward:2005gi}. 

The effective potential for null geodesics is 
\begin{align}
 -\mathcal{U}=\frac{L^2}{r^2(r^3+2M\ell^2)f}\sbrac{\eta r^5-r^3+\brac{\eta A+B}r^2-A}.
\end{align}
Further letting $r=A^{1/3}x$, this is now written as 
\begin{align}
 -\mathcal{U}=-\frac{AL^2}{r^2(r^3+2M\ell^2)f}F(x),
\end{align}
where  
\begin{align}
 F(x)=ax^5+x^3+bx^2+1,\quad a=-\eta\brac{2M\ell^2}^{2/3},\quad b=a-\brac{\frac{2M}{\ell}}^{2/3}. \label{Hayward_F}
\end{align}
As expected from the Class I solution, we can explicitly see how $F(x)$ is a univariate polynomial of four terms, satisfying the conditions of Rojas and Rusek's theorem. This time, the prefactor in front of $F(x)$ is made to be negative so that the normalised coefficients of $F(x)$ have positive signs. With this negative sign in the prefactor, this means stable light rings correspond to $F''(x)>0$ and unstable ones $F''(x)<0$. In particular, the degenerate horizon with $\ell^2=\frac{16}{27}M^2$ is located at $r=\frac{4}{3}M$ and $f''(\frac{4}{3}M)=\frac{9}{8M^2}>0$. Equivalently, $Q''(\frac{4}{3}M)-P''(\frac{4}{3}M)=4M>0$. At $\eta=0$, the function $F(x)$ and $Q-P$ share the same discriminant upon rescaling the coordinate by $r=A^{1/3}x=\brac{2M\ell^2}^{1/3}x$. Therefore $F''(x)>0$ on the degenerate root as well, indicating the presence of the stable light ring.

For general values of $\eta$, the light ring condition $F(x)=F'(x)=0$ leads to 
\begin{align}
 a=-\frac{x^3-2}{3x^5},\quad b=-\frac{2x^3+5}{3x^2}. \label{Hayward_ab}
\end{align}
At $\eta=0$, we have $a=0$ and $b=-\brac{\frac{27}{4}}^{1/3}$.
Following the discriminantal variety \Eqref{Hayward_ab} as $\eta$ continuously increases from zero to positive means $a$ continuously decreases from zero to negative values. The tangent $b'/a'=\frac{\dif b}{\dif a}=-\frac{2(x^3-5)}{x^2}$ being negative means $b$ increases from its $\eta=0$ value. Looking at Eq.~\Eqref{Hayward_F}, to increase in $b$ means $\brac{\frac{2M}{\ell}}^{2/3}$ must decrease from $\brac{\frac{27}{4}}^{1/3}$, taking the parameters into the case of the horizonless Hayward spacetime.

\begin{figure}
 \centering 
 \begin{subfigure}[b]{0.49\textwidth}
    \centering
    \includegraphics[width=\textwidth]{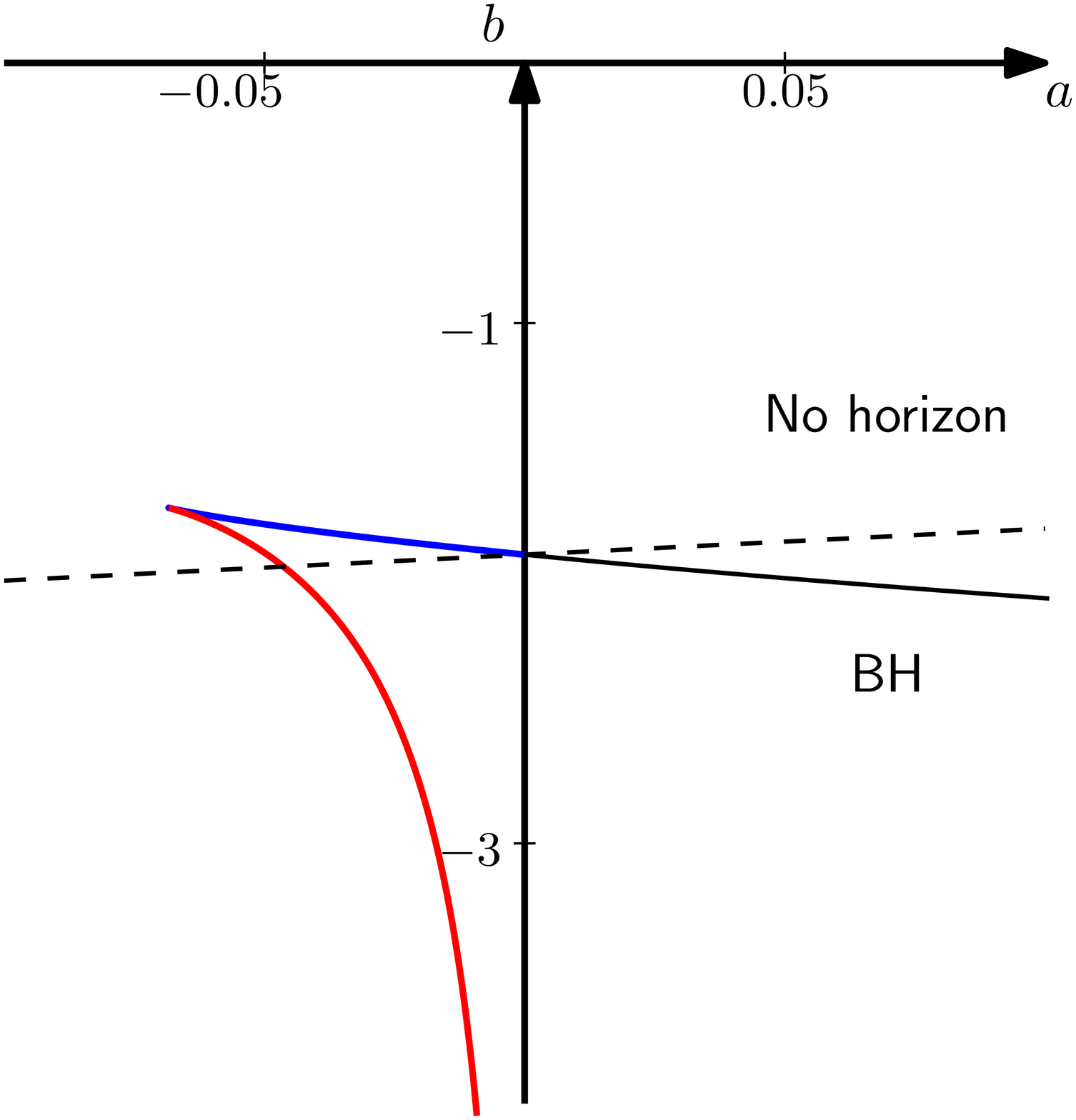}
    \caption{Reduced $A$-discriminant $D_B$.}
    \label{fig_Hay_ab}
  \end{subfigure}
  \begin{subfigure}[b]{0.49\textwidth}
    \centering
    \includegraphics[width=\textwidth]{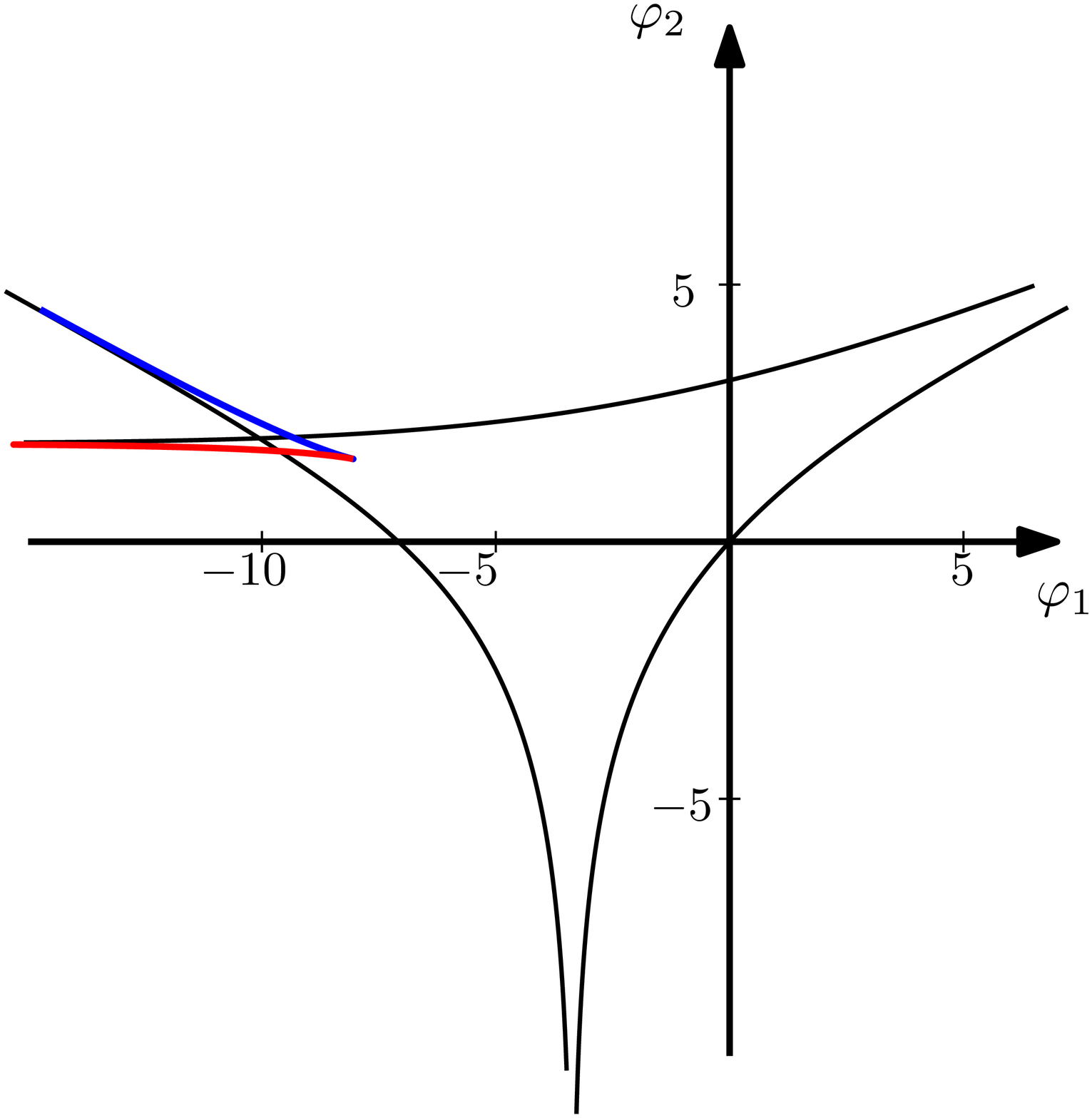}
    \caption{Contour $\mathcal{C}$.}
    \label{fig_Hay_log}
  \end{subfigure}
  \caption{Parameters for light rings in the Hayward spacetime. Fig.~\ref{fig_Hay_ab} shows the reduced $A$-discriminant of $F(x)$ defined in \Eqref{Hayward_F} and \ref{fig_Hay_log} its contour $\mathcal{C}$ of its corresponding amoeba. The domain above the dotted line in the $(a,b)$-plane corresponds to $\frac{4M^2}{\ell^2}>\frac{27}{4}$ of the horizonless Hayward spacetime, and the domain below to $\frac{4M^2}{\ell^2}<\frac{27}{4}$ corresponds to the Hayward black hole. The blue segments of the curves correspond to stable light rings, and the red segments correspond to unstable light rings.}
  \label{fig_Hay}
\end{figure}

The visual depiction is shown in Fig.~\ref{fig_Hay_ab}. The dotted line corresponds to $b=a-\brac{\frac{27}{4}}^{1/3}$. Therefore points above this line correspond to $\frac{4M^2}{\ell^2}>\frac{27}{4}$ for the horizonless Hayward spacetime, and points below correspond to the Hayward black hole with inner and outer horizons. The stable branch of the light ring is shown in blue and lies entirely in the horizonless domain. The red branch is depicted as the red segment which exists in both the horizonless and black hole cases, separted from the stable branch by the cusp.

For completeness, let us obtain the contour corresponding to the discriminantal variety of $F$, as this is where Rojas and Rusek's theorem directly applies. Looking at the exponents of $F(x)$, we see that $A=\{0,2,3,5 \}$. The corresponding $A$-matrix and its Gale dual is
\begin{align}
 A&=\left(\begin{array}{cccc}
                 1 & 1 & 1 & 1 \\
                 0 & 2 & 3 & 5 \\
                \end{array}\right),
                \quad 
         B=\left(\begin{array}{rr}
                2 & -1 \\ 0 & 3 \\ -5 & -2 \\ 3 & 0
               \end{array}\right). \label{Hayward_AB}
\end{align}
The Horn--Kapranov parametrisation is given by
\begin{align}
 \Psi\brac{[\lambda_1:\lambda_2]}=\brac{\frac{27\lambda_1^3(2\lambda_1-\lambda_2)^2}{(-5\lambda_1-2\lambda_2)^5},\;\frac{27\lambda_2^3}{(2\lambda_1-\lambda_2)(-5\lambda_1-2\lambda_2)^2}}.
\end{align}
Taking the restriction of this map to $\mathbb{RP}^{1}$ in the patch where $\lambda_2\neq0$, the components of the map are 
\begin{align}
 \Psi([\lambda:1])&=(\tilde{a},\tilde{b}),\quad \mbox{ where }\quad \tilde{a}=\frac{27\lambda^3(2\lambda-1)^2}{(-5\lambda-2)^5},\quad\tilde{b}=\frac{27}{(2\lambda-1)(-5\lambda-2)^2}.
\end{align}
Finally, the contour $\mathcal{C}$ is the image of $\Psi$ under the log-absolute map,
\begin{align}
 \Phi([\lambda:1])&=\brac{\varphi_1,\varphi_2},\quad\mbox{where}\nonumber\\
 \varphi_1&=\log\left|\frac{27\lambda^3(2\lambda-1)^2}{(-5\lambda-2)^5}\right|,\quad \varphi_2=\log\left|\frac{27}{(2\lambda-1)(-5\lambda-2)^2}\right|.
\end{align}
The image of this map is depicted in Fig.~\ref{fig_Hay_log}. The components $(\tilde{a},\tilde{b})$ are related to $(a,b)$ by 
\begin{align}
 a^3=\tilde{a},\quad b^3=\tilde{b}\quad\leftrightarrow\quad \lambda=\frac{2x^3+5}{x^3-2}.
\end{align}
The exponent of 3 occurs because we have rescaled the entries of $B$ in \Eqref{Hayward_AB} by a factor of 3 to ensure that they are integers.

Rojas and Rusek's theorem in this case tells us that the contour $\mathcal{C}$ contains at most $n=1$ cusps, as indeed demonstrated in Fig.~\ref{fig_Hay_log}. Correspondingly, the discriminantal variety in the $(a,b)$-plane inherits the single cusp, which implies that there exist two branches of light rings. Indeed, the stable branch lies in the domain of horizonless spacetime.

\subsection{The Reissner--Nordstr\"{o}m spacetime}

The Reissner--Nordstr\"{o}m (RN) spacetime is described by the metric 
\begin{subequations}
\begin{align}
 \dif s^2&=-f\dif t^2+f^{-1}\dif r^2+r^2\dif\theta^2+r^2\sin^2\theta\,\dif\phi^2,\\
   f&=1-\frac{2M}{r}+\frac{q^2}{r^2},
\end{align}
\end{subequations}
where $M$ and $q$ are the mass and charge parameters of the spacetime, respectively. It is well known that the solution describes a charged black hole for $|q|<M$. An extremal horizon occurs where $|q|=M$, and the range $|q|>M$ corresponds to the horizonless spacetime with a naked singularity. The stable light rings in the extremal RN spacetime have been studied in \cite{Pradhan:2010ws}. We now reinterpret these results in the context of Rusek's theorem.

In particular, the RN spacetime is an example of a Class IV solution with $A=2M$, $B=-q^2$, $m=2$, $n=1$, and $l=0$. For this case, the effective potential is  
\begin{align}
 -\mathcal{U}&=\frac{L^2}{r^4f}\brac{\eta r^4-r^2+2Mr-q^2}.
\end{align}
Letting $x=-\frac{r}{2m}$, we find 
\begin{align}
 -\mathcal{U}&=-\frac{4M^2L^2}{r^4f}F(x),
\end{align}
where 
\begin{align}
 F(x)=ax^4+x^2+x+b,\quad a=-4M^2\eta,\quad b=\frac{q^2}{4M^2}.
\end{align}
In this case we see that the light-ring condition corresponds to $F=F'=0$, and that stable/unstable light rings correspond to $F(x)$ being a local minimum/maximum, respectively. The light-ring condition leads to
\begin{align}
 a=-\frac{2x+1}{4x^3},\quad b=-\frac{2x^2+3x}{4}. \label{RN_standard}
\end{align}
The point $\eta=0$ corresponds to $a=0$, or $x=-\half$. The value of $b$ at this point is $b=\frac{q^2}{4M^2}=\frac{1}{4}$. This is precisely the extremal condition $|q|=M$. The cusp is located at $a'=b'=0$ which is $x=-\frac{3}{4}$. It can be directly checked that $F''(x)$ is positive between the cusp and $\eta=0$, thus corresponding to stable light rings. The tangent of the curve \Eqref{RN_standard} is given by $b'/a'=\frac{\dif b}{\dif a}=-(4x+3)/x$ which is negative for $-\frac{3}{4}\leq x\leq-\frac{1}{2}$, namely between the cusp and $\eta=0$. Therefore, following the discriminantal variety continuously as $\eta$ increases from zero to positive values is equivalent to following $a$ from zero to negative values. The negative $\frac{\dif b}{\dif a}$ means $b$ must increase as we follow the curve. This will take $\frac{q^2}{4M^2}$ to values greater than $\frac{1}{4}$, thus putting us in the domain of the horizonless RN spacetime with a naked singularity. 

\begin{figure}
 \centering 
 \begin{subfigure}[b]{0.49\textwidth}
    \centering
    \includegraphics[width=\textwidth]{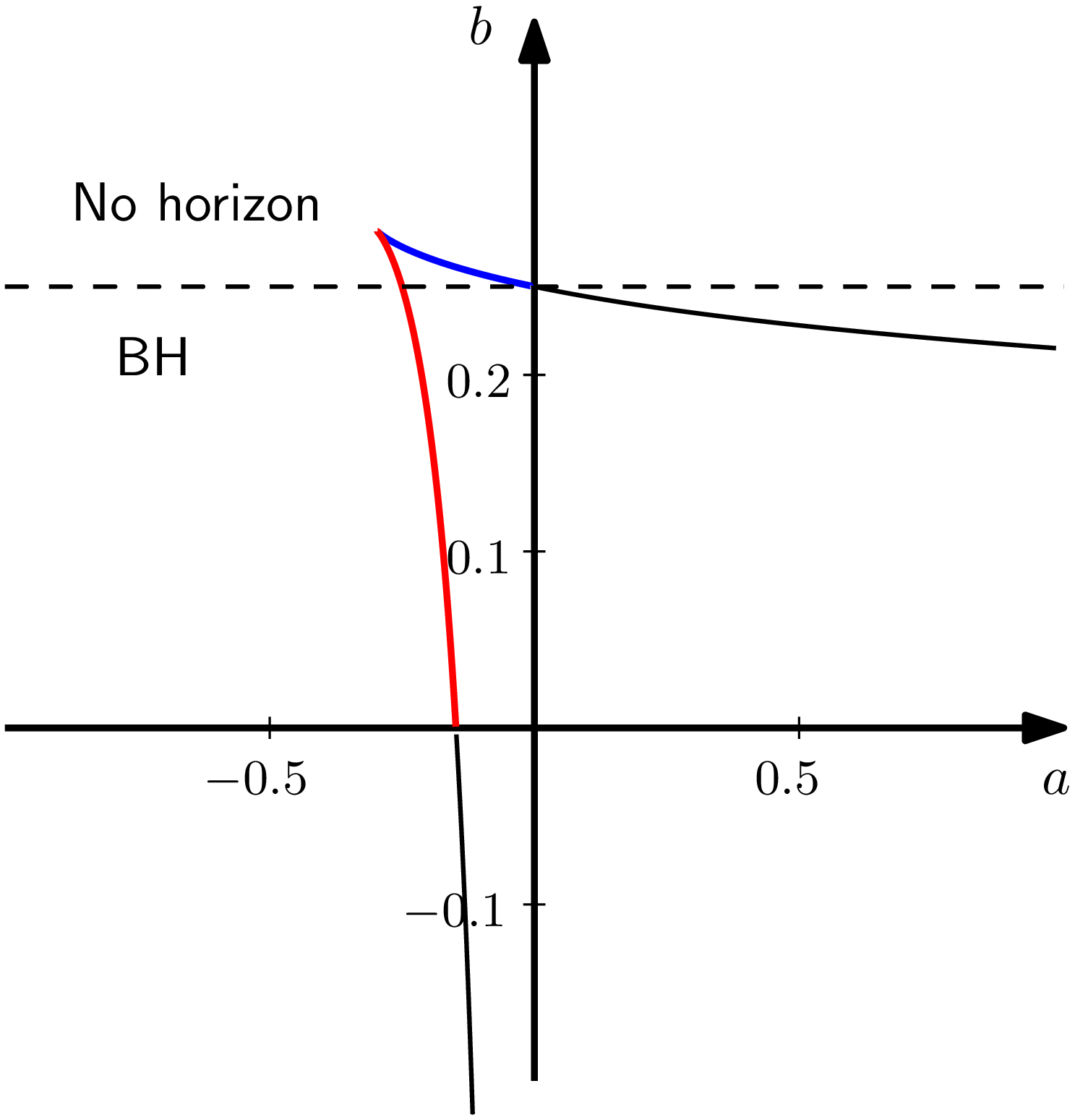}
    \caption{Reduced $A$-discriminant $D_B$.}
    \label{fig_RN_ab}
  \end{subfigure}
  \begin{subfigure}[b]{0.49\textwidth}
    \centering
    \includegraphics[width=\textwidth]{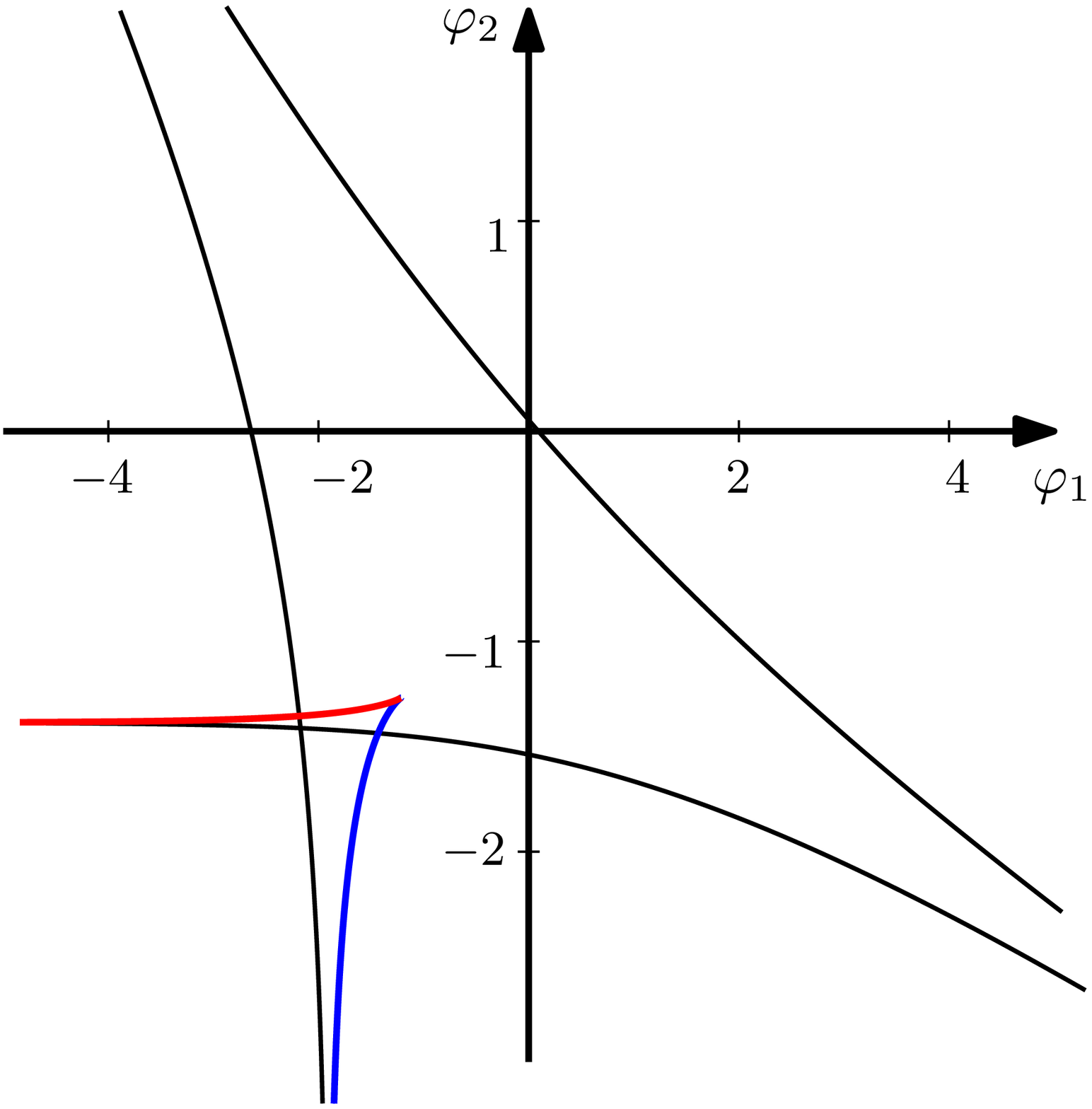}
    \caption{Contour $\mathcal{C}$.}
    \label{fig_RN_log}
  \end{subfigure}
  \caption{The $A$-discriminant and contour $\mathcal{C}$ of the corresponding amoeba for the problem of light rings in the Reissner--Nordstr\"{o}m spacetime. The blue curves correspond to stable light rings, and the red curves correspond to unstable ones. The black curves correspond to light rings with unphysical parameters.}
  \label{fig_RN}
\end{figure}

The preceding discussion is depicted explicitly in the $(a,b)$-plane of Fig.~\ref{fig_RN_ab}, where the horizontal dashed line corresponds to $b=\frac{q^2}{4M^2}=\frac{1}{4}$. Therefore, points above this line correspond to the RN naked singularity, and the points below correspond to the RN black hole. The branch of stable light rings is shown as the blue curve, and the unstable light rings are shown in red. The two branches are separated by the cusp. We see that the stable branch lies entirely in the naked singularity domain. The branches drawn in black are unphysical regimes which require negative $\eta$ or negative $\frac{q^2}{M^2}$.

To complete our discussion, we calculate the contour $\mathcal{C}$ of the discriminantal variety of $F(x)$ where Rojas and Rusek's theorem applies. We see that $A=\{0,1,2,4 \}$, and the corresponding $A$-matrix and its Gale dual is 
\begin{align}
 A=\left(\begin{array}{cccc}
                1 & 1 & 1 & 1 \\
                0 & 1 & 2 & 4 
                \end{array}\right),\quad 
     B=\left(\begin{array}{rr}
          0 & 1 \\ 2 & -2 \\ -3 & 1 \\ 1 & 0
         \end{array}\right).
\end{align}
The map \Eqref{psi_def} gives
\begin{align}
 \Psi\brac{[\lambda_1:\lambda_2]}&=\brac{\frac{\lambda_1(2\lambda_1-2\lambda_2)^2}{(-3\lambda_1+\lambda_2)^3},\,\frac{\lambda_2(-3\lambda_1+\lambda_2)}{(2\lambda_1-2\lambda_2)^2}}.
\end{align}
Taking the patch where $\lambda_1\neq 0$, we have have the $A$-discriminant $D_B$
\begin{align}
 &\Psi\brac{[1:\lambda]}=\brac{a,\,b},\quad\mbox{ where }\quad a=\frac{(2-2\lambda)^2}{-3+\lambda},\quad b=\frac{\lambda(-3+\lambda)}{(2-2\lambda)^2}. \label{RN_HK}
\end{align}
Comparing Eq.~\Eqref{RN_standard} with \Eqref{RN_HK}, we find the two are in agreement if 
\begin{align}
 -x=\frac{\lambda-3}{2(\lambda-1)}=\frac{r}{2m}.
\end{align}
Finally, its contour $\mathcal{C}$ is the image under the log-absolute map,
\begin{align}
 \Phi([1:\lambda])=\brac{\log|a|,\,\log|b|}.
\end{align}
The image of the contour is shown in Fig.~\ref{fig_RN_log}. In particular, we note that the blue (stable light ring) branch in Fig.~\ref{fig_RN_ab} is mapped to the blue branch in Fig.~\ref{fig_RN_log}, and similarly for the red (unstable light ring) branch. Rojas and Rusek's theorem asserts that since $F(x)$ is a univariate ($n=1$) polynomial of $n+3=4$ terms, it has at most $n=1$ cusp, as can be seen in Fig.~\ref{fig_RN_log}. The pre-image of the log-absolute map therefore inherits this single cusp in Fig.~\ref{fig_RN_ab}, where it separates the stable and unstable light ring branches.

\section{Beyond spherical symmetry and \texorpdfstring{$n+3$}{n+3} monomials} \label{sec_beyond}

The examples we considered in the previous section have been for problems involving spherical symmetry, which results in the effective potential being a univariate polynomial, and Rojas and Rusek's theorem applies for the case of 4 monomials. Here, we consider the possibility of applying this approach to more general situations, geodesics in axi-symmetric spacetimes.

\subsection{Example: Light rings around an accelerating black hole}

For an example with two variables, we consider the C-metric, where the metric is 
\begin{subequations}
\begin{align}
 \dif s^2&=\frac{1}{(\tilde{x}-\tilde{y})^2}\brac{-F\dif t^2+\frac{\dif\tilde{y}^2}{F}+\frac{\dif \tilde{x}^2}{G}+G\,\dif\phi^2},\\
   F&=a_0+a_1\tilde{y}+a_2\tilde{y}^2+a_3\tilde{y}^3, \quad G=-a_0-a_1\tilde{x}-a_2\tilde{x}^2+a_3\tilde{x}^3,
\end{align}
\end{subequations}
This solution to the vacuum Einstein equations describe an accelerating black hole. Among the parameters $a_0,a_1, a_2$, and $a_3$, they can be reduced by linear coordinate transformations of $x$ and $y$. It is conventional, e.g., in \cite{Kinnersley:1970zw} to use these transformation to remove the linear terms in $F$ and $G$. However, for the present purposes we instead remove the quadratic term. In other words, we introduce a shift 
\begin{align}
 \tilde{x}\rightarrow x-\frac{a_2}{3a_3},\quad \tilde{y}\rightarrow y-\frac{a_2}{3a_3},
\end{align}
and along with an appropriate renaming of coefficients, the metric becomes\footnote{A third parameter can be fixed to, say, $b_0=1$ by a conformal rescaling of $t$, $x$, $y$, and $\phi$. But it is not necessary for our present purpose.} 
\begin{subequations}\label{C-metric}
 \begin{align}
  \dif s^2&=\frac{1}{(x-y)^2}\brac{-F\dif t^2+\frac{\dif y^2}{F}+\frac{\dif x^2}{G}+G\,\dif\phi^2},\\
   F&=b_0+b_1y+b_3y^3,\quad G=-b_0-b_1x-b_3y^3.
 \end{align}
\end{subequations}
It can be checked that \Eqref{C-metric} still solves the 4-dimensional vacuum Einstein equations, $R_{\mu\nu}=0$.

The effective potential for null geodesics in this spacetime is
\begin{align}
 -\mathcal{U}=\frac{E^2}{F}-\frac{L^2}{G}=\frac{1}{FG}f(x,y),
\end{align}
where 
\begin{align}
 f(x,y)=E^2b_3x^3+L^2b_3y^3+E^2b_1x+L^2b_1y+E^2b_0+L^2b_0.
\end{align}
Circular orbits then correspond to $f=\partial_xf=\partial_yf=0$, which is the defining equations for the discriminant. 

\begin{figure}
 \centering 
 \begin{subfigure}[b]{0.49\textwidth}
    \centering
    \includegraphics[width=0.8\textwidth]{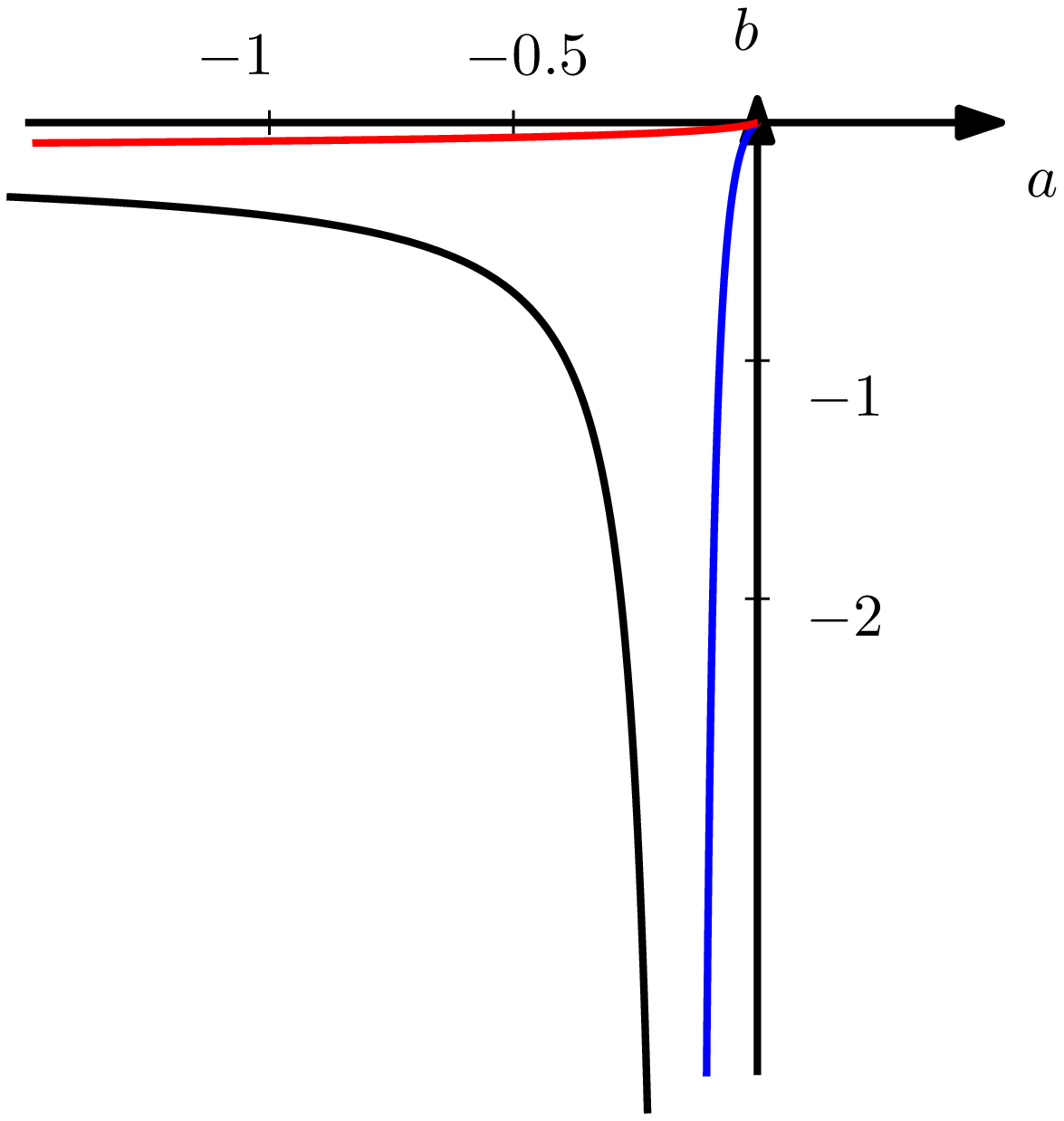}
    \caption{Reduced $A$-discriminant $D_B$.}
    \label{fig_Cmetric_ab}
  \end{subfigure}
  \begin{subfigure}[b]{0.49\textwidth}
    \centering
    \includegraphics[width=0.8\textwidth]{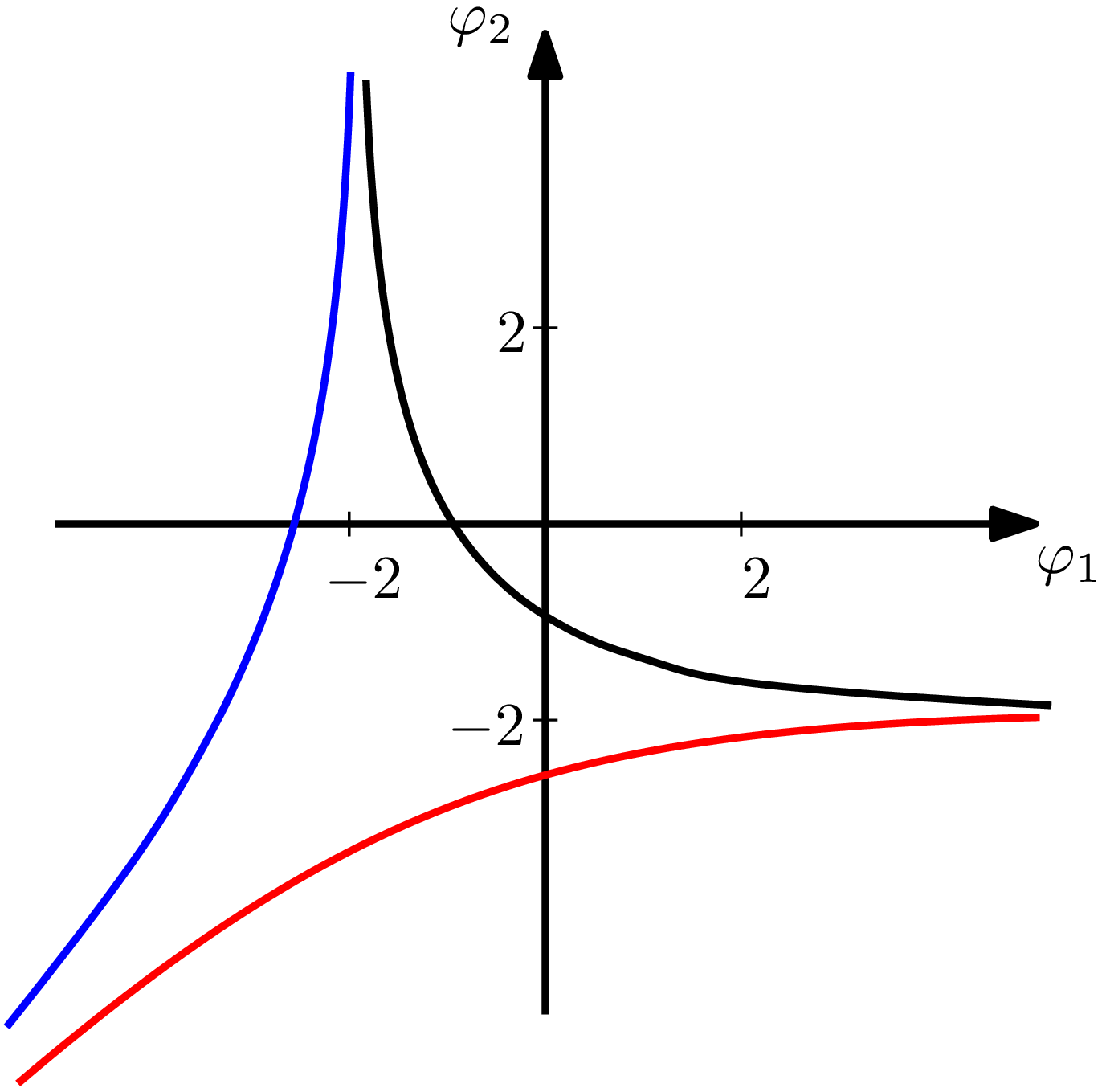}
    \caption{Contour $\mathcal{C}$.}
    \label{fig_Cmetric_log}
  \end{subfigure}
  \caption{The reduced $A$-discriminant $D_B$ for Eq.~\Eqref{Cmetric_ab}, and its corresponding contour $\mathcal{C}$ of its amoeba given by $(\varphi_1,\varphi_2)=\brac{\log|a|,\log|b|}$.}
  \label{fig_Cmetric}
\end{figure}

The $A$-matrix and its corresponding Gale dual is 
\begin{align}
 A=\left(\begin{array}{ccccc}
          1 & 1 & 1 & 1 & 1 \\
          3 & 0 & 1 & 0 & 0 \\
          0 & 3 & 0 & 1 & 0
         \end{array}\right),\quad 
 B=\left(\begin{array}{rr}
          1 & 0 \\
          0 & 1 \\
          -3 & 0\\
          0 & -3 \\
          2 & 2
         \end{array}\right)
\end{align}
Using the Horn--Kapranov procedure, we find 
\begin{align}
 \brac{\frac{b_3b_0^2(E^2+L^2)^2}{E^4b_1^3},\,\frac{b_3b_0^2\brac{E^2+L^2}^2}{L^4b_1^3}} &= \brac{-\frac{4(\lambda+1)^2}{27\lambda^2},\,-\frac{4(\lambda+1)^2}{27}}\nonumber\\
    &=(a,b).\label{Cmetric_ab}
\end{align}
The above parametric curve has a cusp at $\lambda=-1$, or $(0,0)$. Taking the log map, we find that the corresponding amoeba has the cusp at infinity. However $\lambda=-1$ corresponds to either $b_3=0$, $b_0=0$, or $E^2+L^2=0$, all of which are unphysical or irrelevant. It is well known that the C-metric only carries unstable light rings \cite{Pravda:2000zm}. Therefore, there would be no cusps in the physically-relevant parameter ranges of interest. Figure \ref{fig_Cmetric} shows the curve \Eqref{Cmetric_ab} and the contour of its corresponding amoeba.

Eliminating $\lambda$ between the two coordinates above, we recover 
\begin{align}
 27b_3b_0^2E^4+4E^4b_1^3+54b_3b_0E^2L^2-8E^2b_1^3L^2+27b_3b_0^2L^4+4b_1^3L^4=0,
\end{align}
which gives the values of $E$ and $L$ required to satisfy for a circular null geodesic in the C-metric spacetime. This is analogous to the results of \cite{Pravda:2000zm}, although they used a different parametrisation of the C-metric.

\subsection{The case of \texorpdfstring{$n+3$}{n+3} monomials with small additional terms}

In this section, we consider what happens for polynomials of $n+3$ monomials perturbed with small additional terms. We see that the number of cusps may be equal to or less than $n$.

A point on the curve through which more than one branches pass is called a \emph{multiple point}. In particular, a point on a curve through which two branches pass is called a \emph{double point}. A double point at which the two tangents are real and distinct is called a {\em node}, and a double point at which the two tangents are real and coincident is called a {\em cusp}. Finally,  
a double point at which the two tangents are imaginary is called {\em isolated point} or {\em conjugate point}. Let us see that double points on a curve are defined by a function $f(x,y)=0$. First, a double point is a singular point $(x_0,y_0)$ such that 
$\di\frac{\partial f}{\partial x}(x_0,y_0)=\frac{\partial f}{\partial y}(x_0,y_0)=0$. If the determinant of the Hessian is strictly positive, i.e.,
$$
\mathrm{det}\brac{\mathrm{Hess}(f)}\big|_{(x_0,y_0)}=\left(\frac{\partial^2 f}{\partial x\partial y}(x_0,y_0)\right)^2- \frac{\partial^2 f}{\partial x^2}(x_0,y_0)\frac{\partial^2 f}{\partial y^2}(x_0,y_0)>0
$$
then such a double point  $(x_0,y_0)$ is a node. This means that the condition to be a node is an open condition. However, if
$$
\mathrm{det}\brac{\mathrm{Hess}(f)}\big|_{(x_0,y_0)}=\left(\frac{\partial^2 f}{\partial x\partial y}(x_0,y_0)\right)^2- \frac{\partial^2 f}{\partial x^2}(x_0,y_0)\frac{\partial^2 f}{\partial y^2}(x_0,y_0)=0
$$
then such a double point  $(x_0,y_0)$ is a cusp, which implies that being a cusp is a closed condition.

Therefore, the function  which associates to each real algebraic curve $\mathcal{C}$ (real in the sense that the group $\mathbb{Z}_2$ given by the conjugation in $\mathbb{C}$ acts on it, i.e.,  $\mathcal{C} = \bar{\mathcal{C}}$),  the number of its cusps, is  a lower semi-continuous function; i.e., the number of cusps cannot increase if we make a small perturbation of the curve. 

\begin{myex}
 \begin{enumerate}
  \item These are the deformations of the $E_6$ singularities parametrized by $\rho(t)=(t^3, t^4)$, see Fig.~\ref{fig_rho_deformation}.
  \begin{figure}[H]
   \centering 
   \includegraphics{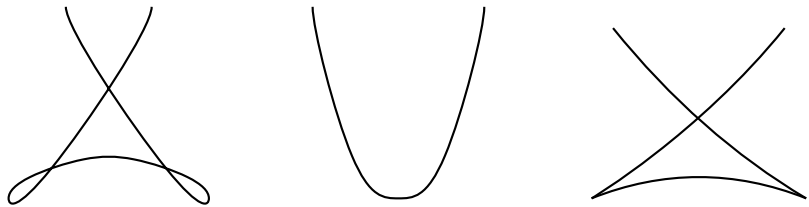}
   \caption{Deformations of the $E_6$ singularities.}
   \label{fig_rho_deformation}
  \end{figure}
  
  \item Another example of simple cubic curves $\mathscr{C}_1$ and $\mathscr{C}_2$ defined by $ y^2=x^3$, and $y^2=x^3 +x$ respectively, is in Fig.~\ref{fig_C1C2}.
  \begin{figure}[H]
   \centering 
   \includegraphics[scale=0.5]{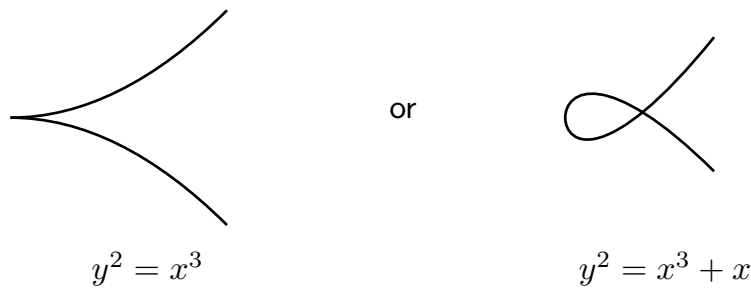}
   \caption{Cubic curves $\mathscr{C}_1$ and $\mathscr{C}_2$.}
   \label{fig_C1C2}
  \end{figure}
  
  \item Other deformations of cusps are such as shown in Fig.~\ref{fig_other_deformations} 
  \begin{figure}[H]
   \centering 
   \includegraphics[scale=0.7]{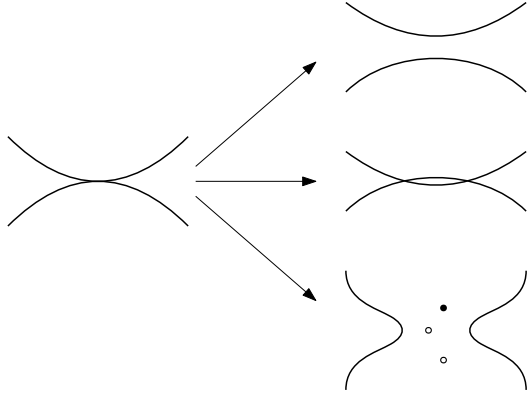}
   \caption{Other deformations of cusps.}
   \label{fig_other_deformations}
  \end{figure}

 \end{enumerate}

\end{myex}

The reduced $A$-discriminant is a polynomial in the coefficients of the polynomial in question. Since being a cusp is a closed condition, as we saw above, then the number of cusps of the amoeba  contour of the reduced $A$-discriminant  $D_{f_\varepsilon}$ of a small perturbation $f_\varepsilon$ of a polynomial $f$ cannot exceed the number of cusps of the amoeba contour of the reduced $A$-discriminant  $D_f$.

There are ways  to remove singularities so not to lose information about the original curve. For one-dimensional curves, it is always possible to carry out this process, called   resolution of the singularity. The  higher-dimensional case  is not so straightforward, but it was done. Oscar Zariski studied singularities in two and three dimensions \cite{Zariski:1939a,Zariski:1939b}, and in 1964, Heisuke Hironaka showed that singularities in any dimension can be resolved \cite{Hironaka:1964a,Hironaka:1964b}. Thus, a polynomial of any degree in any number of variables is equivalent to one that is regular, that is, without singularities. This is not  the subject of our work in this paper. 

Consider a polynomial $f\in (\mathbb{C}^*)^A$   with    $A$ being a finite configuration of points $\{\alpha_{1},\ldots,\alpha_{N} \}\subset\mathbb{Z}^n$ 
\begin{align}
 A=\left(\begin{array}{ccc}
          1 & \cdots & 1 \\
          \alpha_1 & \cdots & \alpha_N
         \end{array}\right)
\label{A_matrix2}
\end{align}
where each $\alpha_i\in\mathbb{Z}^n$ are viewed as column $n$-vectors for $i=1,\ldots,N$. The Gale dual of $A$ is a matrix $B$  with maximal rank and satisfies
\begin{align}
 AB=0. \label{B_def2}
\end{align}
Let $m=N-1-n$, and then the $(N\times m)$-matrix $B$  can be chosen  as follows: 
\begin{align}
 B=\left(\begin{array}{cccc}
          b_1^1 & b_1^2 & \cdots & b_1^m \\
          \vdots & \vdots & \ddots & \vdots \\
          b_{N-m}^1 & b_N^2 & \cdots & b_{N-m}^m\\
          \lambda_1&0&\cdots&0\\
          0&\lambda_2&\cdots&0\\
          \vdots& \vdots & \ddots & \vdots\\
          0&0&\cdots&\lambda_m
         \end{array}\right).
\end{align}

The reduced $A$-discriminant is given by the Horn-Kapranov parametrization corresponding to the matrix $B$. Restricting this parametrization to the real projective space and applying the logarithmic map, we obtain the contour of its amoeba which is a  hypersurface $\mathscr{S}$ in $\mathbb{R}^m$. Now if we remove the $(N+j)$-th column  and the $j$-th row from the matrix $B$ we obtained a parametrization of the reduced $A_j$-discriminant  of the polynomial $g$  obtain by removing the $j$-th monomial from our original polynomial $f$. In other words, the matrix $A_j$ is obtained by removing the $j$-th column from the matrix $A$.  Geometrically, this is the intersection of the hypersurface $\mathscr{S}$ with the $j$-th-plane $H_j$-axis (i.e. $x_j=0$), which is also the $A_j$-discriminant of a polynomial with less monomials than the original one. 
We apply  the same  process to these new $A_j$-discriminant  until we get a discriminant of a polynomial with a lower number of monomials (all this process is done by a routine {\tt maple} calculation), more precisely, until $N-j=n+3$ and then Rojas-Rusek's theorem tells us that the number of cusps is at most $n$ for this last projection. Note that by construction the polynomial corresponding to this last projection has $N-j$ monomials.

\begin{myex}
 Here is another example of univariate polynomial with five monomials. Consider a generic quadric given by  $f = a_0+a_1x+a_2x^2+a_3x^3+a_4x^4$, i.e.,
\begin{align}
 A=\left(\begin{array}{ccccc}
          1 & 1&1&1 & 1 \\
          0 & 1 & 2&3&4 
         \end{array}\right),\quad \textrm{and}
        \quad 
        B=\left(\begin{array}{rrr}
                 3 & 2 & 1 \\
                 -4&-3 & -2 \\
               0& 0&1 \\
                 0&1& 0 \\
                 1&0&0
                \end{array}\right)
\label{A_matrix3}
\end{align}
Therefore, the Horn-Kapranov parametrization is as follows:
\begin{align*}
x_1=\frac{(3+\lambda_2+\lambda_3)^3}{(-4-3\lambda_2-2\lambda_3)^4},\quad x_2=\frac{(3+\lambda_2+\lambda_3)^2\lambda_2}{(-4-3\lambda_2-2\lambda_3)^3}, \quad x_3=\frac{(3+\lambda_2+\lambda_3)\lambda_3}{(-4-3\lambda_2-2\lambda_3)^2}.
\end{align*}
We obtain a surface $\mathscr{S}$ in $\mathbb{R}^3$ such that its intersections with the plane axis  are also  discriminants of polynomials with four monomials coming from the original polynomial $f$ by removing one of its monomials.

The induction on the number of monomials is used as follows.
If we remove the last monomial from $f$, we obtain a new polynomial $f_5$ with four monomials, and  its $A_5$-discriminant is precisely the Horn-Kapranov parametrized given  by removing the last column from $A$, and $B_5$ is obtained by removing   the first column and the fifth row from the matrix $B$ to obtain the following:
\begin{align}
 A_5=\left(\begin{array}{cccc}
          1 & 1&1&1  \\
          0 & 1 & 2&3
         \end{array}\right),\quad \textrm{and}
        \quad 
        B_5=\left(\begin{array}{cc}
                  2 & 1 \\
                 -3 & -2 \\
               0&1 \\
                 1& 0 
                 \end{array}\right).
\end{align}
If we remove, for example, the third monomial from $f$, we obtain a new polynomial $f_3$, and  its $A_3$-discriminant is precisely the Horn-Kapranov parametrized given  by removing the third column from $A$,  and $B_3$ is obtained by removing   the third column and the third  row from the matrix $B$ to obtain the following: 
\begin{align}
 A_3=\left(\begin{array}{cccc}
          1 & 1&1&1  \\
          0 & 1 &3&4 
         \end{array}\right),\quad \textrm{and}
        \quad
        B_3=\left(\begin{array}{rr}
                 3 & 2  \\
                 -4&-3  \\ 
                 0&1 \\
                 1&0
                \end{array}\right).
\end{align}
We presented  here two projections, and the third one can be done in the same way.
\end{myex}

\section{Conclusion} \label{sec_conclusion} 

In this paper, we have applied the theory of $A$-discriminantal varieties to study time-like and null circular geodesics of various spacetimes. In particular, Rojas and Rusek's theorem is applied to identify classes of algebraic spacetimes with at most two branches of light rings, one of which is stable and the other unstable. It was also shown that the unstable branch always occurs in parameters for which the spacetime is horizonless. 

Perhaps it is worth noting that the conditions for which Rojas and Rusek's theorem applies is highly stringent. First, the functions involved must be algebraic, and second, the resulting effective potential must take the form of a univariate polynomial with exactly four terms. The fact that the functions must be algebraic immediately makes it non-applicable to spacetimes with functions of non-integer exponents. For instance, the $g_{tt}$ component of the Fisher/JNW spacetime \cite{Fisher:1948yn,Janis:1968zz} takes the form $-(1-r_0/r)^\nu$ and is non-algebraic since $0<\nu<1$ for this solution. Remarkably, even with these restrictions, we still have many examples of spacetimes that \emph{do} satisfy the conditions. The examples provided in this paper have been time-like geodesics around the Schwarzschild black hole, light rings in the Hayward, Reissner--Nordstr\"{o}m, and C-metric solutions.

In any case, for the `algebraic' spacetimes we have identified as Class I--IV that light rings do indeed come in pairs, thus giving support to the results of \cite{Cunha:2017qtt,Cunha:2020azh,Guo:2020qwk} from the perspective of $A$-discriminants. All of these appear to show that unstable light rings seem to be a generic feature of spherically symmetric, asymptotically flat spacetimes.

Our results makes use of a specific result in the form of Rojas and Rusek's theorem. At the same time, there is a vast literature on discriminants and varieties, and many of these are reviewed in \cite{GKZ-94} and other references. It may be worth exploring further to see how the present results can be expanded and/or generalised.

\section*{Acknowledgements}

M.~N. is supported by Xiamen University Malaysia Research Fund (Grant No. XMUMRF/2020-C5/IMAT/0013). 
Y.-K.~L. is supported by Xiamen University Malaysia Research Fund (Grant No. XMUMRF/2021-C8/IPHY/0001). 

\bibliographystyle{Adiscriminant-geod}

\bibliography{Adiscriminant-geod}

\end{document}